\PassOptionsToPackage{hyphens}{url}
\documentclass[letterpaper,twocolumn,10pt]{article}
\usepackage{usenix-2020-09}

\usepackage{hyperref}
\usepackage{xspace}
\usepackage{multirow}
\usepackage{multicol} 
\usepackage{comment}
\usepackage{enumerate}
\usepackage{graphicx}
\usepackage{tikz}
\usepackage{wasysym}
\usepackage{booktabs}
\usepackage{balance} 
\usepackage[inline]{enumitem} 
\usepackage{tabularx}
\usepackage{csquotes}
\usepackage{siunitx}
\usepackage{caption}
\usepackage{subcaption}
\usepackage{cleveref}
\usepackage{lscape}

\usepackage[all=normal,bibbreaks=tight]{savetrees}

\usepackage{titlesec}
 \titlespacing*{\section}{0pt}{1ex}{.7ex}
 \titlespacing{\subsection}{0pt}{1ex}{.7ex}
 \titlespacing{\paragraph}{0pt}{1.5ex}{1.5ex}
 
\DeclareSIUnit[round-mode = places, round-precision = 0]{\percent}{\%}
\DeclareSIUnit[number-unit-product = \,]{\USD}{\textsc{USD}}
\DeclareSIUnit{\dollar}{\$}
\DeclareSIPrefix{\million}{\text{ million }}{6}
\DeclareSIPrefix{\billion}{\text{ billion }}{9}

\hypersetup{%
  pdflang={en-US},
  pdfkeywords={privacy; user study},
  pdfdisplaydoctitle=true, 
  bookmarksnumbered,
  pdfborder = {0 0 0},
  colorlinks,
  linkcolor={black!80!black},
  citecolor={red!70!black},
  urlcolor={blue!70!black},
  pdfstartview={FitH},
  breaklinks=true,
  hypertexnames=false
}

\newcommand{\changes}[1]{{\color{black}{#1}}} 

\newcommand{\ie}{i.\,e.}

\newcommand{\etal}{et~al.\@\,}

\newcolumntype{L}[1]{>{\hsize=#1\hsize\raggedright\arraybackslash}X}%
\newcolumntype{R}[1]{>{\hsize=#1\hsize\raggedleft\arraybackslash}X}%
\newcolumntype{C}[1]{>{\hsize=#1\hsize\centering\arraybackslash}X}%

\newcolumntype{H}{>{\iffalse}c<{\fi}@{}}

\newlist{compactitem}{itemize}{5}
\setlist[compactitem]{leftmargin=*, nosep}
\setlist[compactitem, 1]{label=\textbullet}
\setlist[compactitem, 2]{label=\textendash}
\setlist[compactitem, 3]{label=\textasteriskcentered}
\setlist[compactitem, 4]{label=\textperiodcentered}

\makeatletter
\let\orgItem\item
\NewDocumentCommand\fixedItem{ o }{%
   \IfNoValueTF{#1}%
      {\orgItem}
      {\orgItem[#1]\def\@currentlabel{#1}}
}
\makeatother

\newlist{questions}{enumerate}{3}
\setlist[questions]{align=left, labelwidth=2em, labelsep=.5em, listparindent=0pt, itemindent=0pt, leftmargin=!, before=\let\item\fixedItem}
\setlist[questions, 1]{labelindent=0pt, label=\textbf{Q\arabic*}, widest=99}
\setlist[questions, 2]{labelindent=-2.5em, label*=\textbf{\_\Alph*}, widest=26}
\setlist[questions, 3]{labelindent=-2.5em, label*=\textbf{\_\roman*}, widest=9}

\newlist{answers}{itemize}{1}
\setlist[answers]{leftmargin=*, nosep, align=left, label=$\bigcirc$}
\newlist{answers*}{itemize*}{1}
\setlist[answers*]{label=$\bigcirc$}

\microtypecontext{spacing=nonfrench}

\usepackage{titlesec}

\titlespacing{\subparagraph}{0pt}{1.9pt}{1.5ex}
\titleformat*{\subparagraph}{\normalsize\em}

\begin{document}

\date{}

\title{\Large Security and Privacy Perceptions of \\Third-Party Application Access for Google Accounts \\ (Extended Version)}

\def\plainauthor{Author name(s) for PDF metadata. Don't forget to anonymize for submission!}

\author{
{\rm David G. Balash}\\
The George Washington University
\and
{\rm Xiaoyuan Wu}\\
The George Washington University
\and
{\rm Miles Grant}\\
The George Washington University
\and
{\rm Irwin Reyes}\\
Two Six Technologies
\and
{\rm Adam J. Aviv}\\
The George Washington University
} 

\maketitle


\newcommand{\tabDemographicsAndIUIPC}[0]{
\begin{table}[t]
\centering
\small
\caption[Characteristics of the study participants]{\label{tab:characteristics}
Demographic and IUIPC data collected at the end of the first survey.
}
\renewcommand{\arraystretch}{0.6}
\begin{tabular*}{\columnwidth}{@{}>{\bfseries}ll@{\extracolsep{\fill}}*{4}{r}}
\toprule
&                & \multicolumn{2}{c}{\textbf{First}} & \multicolumn{2}{c}{\textbf{Second}} \\
&                & \multicolumn{2}{c}{\textbf{Survey}} & \multicolumn{2}{c}{\textbf{Survey}} \\
&                & \multicolumn{2}{c}{\emph{(n = \num{432})}} & \multicolumn{2}{c}{\emph{(n = \num{214})}} \\
\midrule
\multirow{6}{*}{\rotatebox[origin=c]{90}{Gender}}
& & \multicolumn{1}{c}{\textbf{n}} & \multicolumn{1}{c}{\textbf{\%}} & \multicolumn{1}{c}{\textbf{n}} & \multicolumn{1}{c}{\textbf{\%}} \\
\cmidrule{3-6}
%

%
& Woman          & 204    &  47     &   99    &  46 \\
& Man            & 216    &  50     &  107    &  50 \\
& Non-binary     &  11    &   3     &    7    &   3 \\
& No answer      &   1    &   0     &    1    &   1 \\
\midrule
\multirow{6.5}{*}{\rotatebox[origin=c]{90}{Age}}
%
%
%
& 18--24         & 132    & 31  &  66    & 31  \\
& 25--34         & 157    & 36  &  80    & 37  \\
& 35--44         &  75    & 17  &  32    & 15  \\
& 45--54         &  44    & 10  &  20    &  9  \\
& 55+            &  23    &  6  &  15    &  7  \\
& No answer      &   1    &  0  &   1    &  1  \\
\midrule\multirow{7}{*}{\rotatebox[origin=c]{90}{IUIPC}}
& & \multicolumn{1}{c}{\textbf{Avg.}} & \multicolumn{1}{c}{\textbf{SD}} & \multicolumn{1}{c}{\textbf{Avg.}} & \multicolumn{1}{c}{\textbf{SD}} \\
\cmidrule{3-6}
%
%

& Control        &   5.9 &  1.0 &   5.8 &  1.0 \\
& Awareness      &   6.6 &  0.7 &   6.6 &  0.8 \\
& Collection     &   5.4 &  1.2 &   5.3 &  1.2 \\
\cmidrule{2-6}
& IUIPC Combined &   5.8 &  0.8 &   5.8 &  0.8 \\
\bottomrule
\end{tabular*}
\end{table}

}

\newcommand{\tabDemographicsFull}[0]{
\begin{table}[ht]
    \caption{Full demographics data of the participants of the first survey. \label{tab:demographicsFull}}
    \centering
    \begin{tabular}{r|rrrrrrrr|rr}
    \hline
    & \multicolumn{2}{c}{\textbf{Male}} & \multicolumn{2}{c}{\textbf{Female}}
    & \multicolumn{2}{c}{\textbf{Non-binary}} & \multicolumn{2}{c|}{\textbf{PND}} & \multicolumn{2}{c}{\textbf{Total}} \\
    & \textbf{No.} & \textbf{\%} & \textbf{No.} & \textbf{\%}
    & \textbf{No.} & \textbf{\%} & \textbf{No.} & \textbf{\%} & \textbf{No.} & \textbf{\%}\\
    \hline
    \textbf{Age} & 216 & 50 & 205 & 47 & 11 & 3 & 2 & 0 & 434 & 100\\
    \hline
    18 - 24 & 58 & 13 & 71 & 16 & 4 & 1 & 0 & 0 & 133 & 31\\
    25 - 34 & 84 & 19 & 67 & 15 & 6 & 1 & 0 & 0 & 157 & 36\\
    35 - 44 & 44 & 10 & 30 & 7 & 1 & 0 & 0 & 0 & 75 & 17\\
    45 - 54 & 19 & 4 & 25 & 6 & 0 & 0 & 0 & 0 & 44 & 10\\
    55 - 64 & 7 & 2 & 8 & 2 & 0 & 0 & 0 & 0 & 15 & 3\\
    65 or older & 4 & 1 & 4 & 1 & 0 & 0 & 0 & 0 & 8 & 2\\
    Prefer not to disclose & 0 & 0 & 0 & 0 & 0 & 0 & 2 & 0 & 2 & 0\\
    \hline
    \textbf{Highest level of school} & 216 & 50 & 205 & 47 & 11 & 3 & 2 & 0 & 434 & 100\\
    \hline
    No schooling completed & 0 & 0 & 0 & 0 & 0 & 0 & 0 & 0 & 0 & 0\\
    Some high school & 3 & 1 & 1 & 0 & 0 & 0 & 0 & 0 & 4 & 1\\
    High school & 20 & 5 & 17 & 4 & 2 & 0 & 0 & 0 & 39 & 9\\
    Some college & 48 & 11 & 37 & 9 & 4 & 1 & 0 & 0 & 89 & 21\\
    Trade & 4 & 1 & 5 & 1 & 0 & 0 & 0 & 0 & 9 & 2\\
    Associate's Degree & 19 & 4 & 7 & 2 & 1 & 0 & 0 & 0 & 27 & 6\\
    Bachelor's Degree & 89 & 21 & 91 & 21 & 3 & 1 & 0 & 0 & 183 & 42\\
    Master's Degree & 24 & 6 & 36 & 8 & 1 & 0 & 0 & 0 & 61 & 14\\
    Professional degree & 4 & 1 & 4 & 1 & 0 & 0 & 0 & 0 & 8 & 2\\
    Doctorate & 5 & 1 & 6 & 1 & 0 & 0 & 0 & 0 & 11 & 3\\
    Prefer not to disclose & 0 & 0 & 1 & 0 & 0 & 0 & 2 & 0 & 3 & 1\\
    \hline
    \textbf{Background} & 216 & 50 & 205 & 47 & 11 & 3 & 2 & 0 & 434 & 100\\
    \hline
    Technical & 74 & 17 & 25 & 6 & 1 & 0 & 0 & 0 & 100 & 23\\
    Non-Technical & 134 & 31 & 178 & 41 & 10 & 2 & 0 & 0 & 322 & 74\\
    Prefer not to disclose & 8 & 2 & 2 & 0 & 0 & 0 & 2 & 0 & 12 & 3\\
    \end{tabular}
\end{table}
}

\newcommand{\tabDemographicsMain}[0]{
\begin{table}[ht]
    \caption{Full demographics data of the participants of the second survey. \label{tab:demographicsMainStudy}}
    \centering
    \begin{tabular}{r|rrrrrrrr|rr}
    \hline
    & \multicolumn{2}{c}{\textbf{Male}} & \multicolumn{2}{c}{\textbf{Female}} 
    & \multicolumn{2}{c}{\textbf{Non-binary}} & \multicolumn{2}{c|}{\textbf{PND}} & \multicolumn{2}{c}{\textbf{Total}} \\
    & \textbf{No.} & \textbf{\%} & \textbf{No.} & \textbf{\%}
    & \textbf{No.} & \textbf{\%} & \textbf{No.} & \textbf{\%} & \textbf{No.} & \textbf{\%}\\
    \hline
    \textbf{Age} & 68 & 46 & 75 & 51 & 4 & 3 & 1 & 1 & 148 & 100\\
    \hline
    18 - 24 & 18 & 12 & 34 & 23 & 3 & 2 & 0 & 0 & 55 & 37\\
    25 - 34 & 24 & 16 & 21 & 14 & 1 & 1 & 0 & 0 & 46 & 31\\
    35 - 44 & 15 & 10 & 9 & 6 & 0 & 0 & 0 & 0 & 24 & 16\\
    45 - 54 & 7 & 5 & 7 & 5 & 0 & 0 & 0 & 0 & 14 & 9\\
    55 - 64 & 2 & 1 & 3 & 2 & 0 & 0 & 0 & 0 & 5 & 3\\
    65 or older & 2 & 1 & 1 & 1 & 0 & 0 & 0 & 0 & 3 & 2\\
    Prefer not to disclose & 0 & 0 & 0 & 0 & 0 & 0 & 1 & 1 & 1 & 1\\
    \hline
    \textbf{Highest level of school} & 68 & 46 & 75 & 51 & 4 & 3 & 1 & 1 & 148 & 100\\
    \hline
    No schooling completed & 0 & 0 & 0 & 0 & 0 & 0 & 0 & 0 & 0 & 0\\
    Some high school & 0 & 0 & 0 & 0 & 0 & 0 & 0 & 0 & 0 & 0\\
    High school & 7 & 5 & 3 & 2 & 1 & 1 & 0 & 0 & 11 & 7\\
    Some college & 20 & 14 & 13 & 9 & 2 & 1 & 0 & 0 & 35 & 24\\
    Trade & 1 & 1 & 2 & 1 & 0 & 0 & 0 & 0 & 3 & 2\\
    Associate's Degree & 2 & 1 & 1 & 1 & 0 & 0 & 0 & 0 & 3 & 2\\
    Bachelor's Degree & 28 & 19 & 40 & 27 & 1 & 1 & 0 & 0 & 69 & 47\\
    Master's Degree & 8 & 5 & 13 & 9 & 0 & 0 & 0 & 0 & 21 & 14\\
    Professional degree & 1 & 1 & 1 & 1 & 0 & 0 & 0 & 0 & 2 & 1\\
    Doctorate & 1 & 1 & 1 & 1 & 0 & 0 & 0 & 0 & 2 & 1\\
    Prefer not to disclose & 0 & 0 & 1 & 1 & 0 & 0 & 1 & 1 & 2 & 1\\
    \hline
    \textbf{Background} & 68 & 46 & 75 & 51 & 4 & 3 & 1 & 1 & 148 & 100\\
    \hline
    Technical & 21 & 14 & 11 & 7 & 0 & 0 & 0 & 0 & 32 & 22\\
    Non-Technical & 44 & 30 & 64 & 43 & 4 & 3 & 0 & 0 & 112 & 76\\
    Prefer not to disclose & 3 & 2 & 0 & 0 & 0 & 0 & 1 & 1 & 4 & 3\\
    \end{tabular}
\end{table}
}

\newcommand{\tabRemoveAppRegression}[0]{
\begin{table}[tbp]
\centering
\caption[Remove app regression analysis]{\label{tab:remove-app-regression}
Binomial logistic model to describe which factors influenced the preference to remove an app (\emph{Remove} responses to question \ref{appendix:main-survey:Q5}).
The  Aldrich-Nelson pseudo  $R^2 = \num{0.52}$.
}
{
\small
\begin{tabular*}{\columnwidth}{
@{}
l 
@{\extracolsep{\fill}}
S[table-format = -1.2]
S[table-format = 1.2]
@{\extracolsep{4pt}}
S[table-format = 1.2, parse-numbers = false, parse-units = false]
@{\extracolsep{4pt}}H@{\extracolsep{\fill}}
S[table-format = <1.3]
@{\extracolsep{1pt}}
l
}
    \toprule
    {\textbf{Factor}} & {\textbf{Est.}} & {\textbf{OR}} & {\textbf{Error}} & {\textbf{z value}} & {\textbf{Pr(\textgreater\textbar z\textbar)}} & \\ 
    \midrule
(Intercept) & -1.02 & 0.36 & 0.68 & -1.50 & 0.134 &  \\ 
$\text{Participant's newest app}$ & 0.22 & 1.24 & 0.53 & 0.40 & 0.687 &  \\
$\text{Participant's oldest app}$ & -0.07 & 0.93 & 0.56 & -0.12 & 0.904 &  \\ 
$\text{Recall} = Yes$ & 0.92 & 2.50 & 0.68 & 1.34 & 0.179 &  \\ 
$\text{Aware app permissions} = Yes$ & -1.15 & 0.32 & 0.61 & -1.90 & 0.058 & . \\ 
$\text{Last use} \in \{\textit{day, week, month}\}$ & -1.76 & 0.17 & 0.49 & -3.57 & <0.001 & *** \\ 
$\text{Access benef.} \in \{\textit{Agr., Str. Ag.}\}$ & -1.80 & 0.17 & 0.55 & -3.29 & 0.001 & ** \\ 
$\text{Access conc.} \in \{\textit{Agr., Str. Ag.}\}$ & 0.43 & 1.53 & 0.56 & 0.77 & 0.443 &  \\ 
$\text{Access change} \in \{\textit{Agr., Str. Ag.}\}$ & 1.75 & 5.76 & 0.54 & 3.25 & 0.001 & ** \\ 
$\text{\# of permissions} > \text{median}$ & 0.05 & 1.05 & 0.44 & 0.11 & 0.909 &  \\ 
$\text{Time since install} < \text{3 months}$ & 0.84 & 2.32 & 0.54 & 1.54 & 0.123 &  \\
$\text{Time since install} > \text{2 years}$ & 0.26 & 1.30 & 0.66 & 0.40 & 0.689 &  \\
  \bottomrule
\end{tabular*}
}
\footnotesize
\textbf{Signif. codes:} $\text{***}~\widehat{=} < 0.001$; $\text{**}~\widehat{=} <0.01$; $\text{*}~\widehat{=} <0.05$; $\cdot~\widehat{=} <0.1$
\end{table}

}

\newcommand{\tabAppsPermissionsTable}[0]{
\renewcommand{\arraystretch}{1.25}
\begin{table*}[h]
\caption{The top 28 (5-way tie for permissions count of 8) most requested permissions by third-party apps with authorized access to participants' Google accounts, the permission category, and a count of how many of each permissions were authorized.\label{tab:app-permissions}}
\footnotesize \centering
\begin{tabular}{r|p{2.3cm}|l}
     \textbf{Count} & \textbf{Category}     & \textbf{Permission} \\
     \toprule
     223            & Google Play           & Create, edit, and delete your Google Play Games activity \\
     189            & Basic account info    & See your primary Google Account email address \\
     177            & Basic account info    & See your personal info, including any personal info you've made publicly available \\
     71             & Google Drive          & See, create, and delete its own configuration data in your Google Drive \\
     44             & Basic account info    & Associate you with your personal info on Google \\
     28             & Google Drive          & See, edit, create, and delete only the specific Google Drive files you use with this app \\
     27             & Google Drive          & See, edit, create, and delete all of your Google Drive files \\
     27             & Gmail                 & Read, compose, send, and permanently delete all your email from Gmail \\
     26             & Google Contacts       & See, edit, download, and permanently delete your contacts \\
     24             & Google Calendar       & See, edit, share, and permanently delete all the calendars you can access using Google Calendar \\
     18             & Google Contacts       & See and download your contacts \\
     16             & Basic account info    & Full account access \\
     14             & Additional access     & See and download your exact date of birth \\
     13             & Additional access / Basic account info    & Display and run third-party web content in prompts and sidebars inside Google applications \\
     12             & Gmail                 & View your email messages and settings \\
     12             & Google Drive          & See and download all your Google Drive files \\
     11             & Additional access     & See your age group \\
     11             & Additional access / Basic account info     & Use Google Fit to see and store your physical activity data \\
     10             & Google Docs           & See, edit, create, and delete your spreadsheets in Google Drive \\
     10             & Google Drive          & Connect itself to your Google Drive \\
     10             & Additional access     & Connect to an external service \\
     9              & Additional access     & See your language preferences \\
     9              & Additional access / Basic account info   & See and add to your Google Fit physical activity data \\
     8              & Additional access / Basic account info   & See and add info about your body measurements and heart rate to Google Fit \\
     8              & Gmail                 & Send email on your behalf\\
     8              & Google Docs           & See, create, and edit all Google Docs documents you have access to\\
     8              & Additional access     & See and download all of your Google Account email addresses\\
     8              & YouTube               & Manage your YouTube account\\
\end{tabular}
\end{table*}

}

\newcommand{\tabSSOsPermissionsTable}[0]{
\begin{table*}[h]
\caption{The top 25 most requested permissions by SSO with authorized access to participants' Google accounts ranked by the number of each permission authorized, and the permission category.\label{tab:sso-permissions}}
\footnotesize \centering
\begin{tabular}{r|l|l}
     \textbf{Count} & \textbf{Category} & \textbf{Permission} \\
     \toprule
     976            & Basic account info    & See your primary Google Account email address \\
     938            & Basic account info    & See your personal info, including any personal info you've made publicly available \\
     88             & Basic account info    & Associate you with your personal info on Google \\
     45             & Google Play           & Create, edit, and delete your Google Play Games activity \\
     27             & Google Drive          & See, edit, create, and delete only the specific Google Drive files you use with this app \\
     25             & Gmail                 & Read, compose, send, and permanently delete all your email from Gmail \\
     25             & Google Calendar       & See, edit, share, and permanently delete all the calendars you can access using Google Calendar \\
     24             & Google Contacts       & See, edit, download, and permanently delete your contacts \\
     20             & Google Contacts       & See and download your contacts \\
     20             & Google Drive          & See, edit, create, and delete all of your Google Drive files \\
     16             & Google Drive          & See, create, and delete its own configuration data in your Google Drive \\
     14             & Additional access     & See and download your exact date of birth \\
     12             & Gmail                 & View your email messages and settings \\
     11             & Additional access     & See your age group \\
     11             & Google Drive          & See and download all your Google Drive files \\
     10             & Google Docs           & See, edit, create, and delete your spreadsheets in Google Drive \\
     10             & Google Drive          & Connect itself to your Google Drive \\
     10             & Additional access     & Display and run third-party web content in prompts and sidebars inside Google applications \\
     9              & Additional access     & See your language preferences \\
     9              & Gmail                 & Send email on your behalf \\
     9              & Additional access     & See and add info about your body measurements and heart rate to Google Fit \\
     9              & Additional access     & See and add to your Google Fit physical activity data \\
     9              & Additional access     & Connect to an external service \\
     8              & Additional access     & See and download all of your Google Account email addresses \\
     8              & Additional access     & Use Google Fit to see and store your physical activity data \\
\end{tabular}
\end{table*}
}

\newcommand{\tabAppsTable}[0]{

\begin{table*}[h]
    \caption{The top 26 (4 apps tied for 23rd ranked by authorized count) apps with access, the categories for the requested permissions, the average number of days authorized to the participants' Google account, and the number of permissions each app requested.\label{tab:apps}}
    \footnotesize \centering
    \begin{tabular}{r|l|c|c|c}
    
                                                    &                                   & \textbf{Number}  & \textbf{Avg Num of}       & \textbf{Number of}    \\
        \textbf{App}                           & \textbf{Permission Categories}    &\textbf{Authorized}     & \textbf{Days Authorized}   & \textbf{Permissions}  \\
        \toprule
        Dropbox                                     & Basic account info, Google Contacts & 40                & 502                       & 4                     \\
        SAMSUNG Account                             & Additional access, Basic account info & 38                & 63                        & 5                     \\
        Shop: package \& order tracker              & Basic account info, Gmail         & 32                & 37                        & 3                     \\
        Nest                                        & Additional access, Basic account info & 31                & 265                       & 3                     \\
        Windows                                     & Basic account info, Gmail, Google Calendar & 29                & 397                       & 8                     \\
                                                    & Google Contacts                   &                   &                           &                       \\
        Microsoft SwiftKey Keyboard                 & Additional access, Basic account info & 27                & 102                       & 3                     \\
        macOS                                       & Basic account info, Gmail, Google Calendar, & 19                & 452                       & 6                     \\
                                                    & Google Contacts, Google Hangouts    &                   &                           &                       \\
        Zoom                                        & Basic account info, Google Calendar,  & 16                & 158                       & 5         \\
                                                    & Google Contacts                       &                   &                           &           \\
        WhatsApp Messenger                          & Google Drive                      & 13                & 663                       & 2                     \\
        Rakuten Cash Back                           & Basic account info, Gmail         & 13                & 268                       & 3                     \\
        Pokémon GO                                  & Additional access, Basic account info & 10                & 307                       & 4                     \\
        Microsoft apps \& services                  & Additional access, Basic account info, Gmail, & 10            & 290                       & 8                     \\
                                                    & Google Calendar, Google Contacts, Google Drive &                       &                       \\
        Samsung Email                               & Basic account info, Gmail         & 9                 & 1,014                    & 4                     \\
        Paribus                                     & Basic account info, Gmail         & 9                 & 390                       & 8                     \\
        Unroll.me                                   & Basic account info, Gmail, Google Contacts & 8                 & 1,389                    & 5                     \\
        Clash Royale                                & Google Play                       & 8                 & 644                       & 1                     \\
        Google Nest Hub                             & Basic account info                & 8                 & 287                       & 1                     \\
        PlayStation Network                         & Basic account info, YouTube       & 7                 & 573                       & 5                     \\
        Shop                                        & Basic account info, Gmail         & 7                 & 441                       & 5                     \\
        Chromecast                                  & Basic account info                & 7                 & 225                       & 1                     \\
        Lumin PDF                                   & Basic account info, Google Drive  & 7                 & 164                       & 5                     \\
        DocHub - PDF Sign \& Edit                   & Basic account info, Gmail, Google Contacts & 7                 & 86                        & 6                     \\
                                                    & Google Drive                      &                   &                           &                       \\
        IFTTT                                       & Basic account info                & 6                 & 1,130                    & 1                     \\
        Sweatcoin — Walking step counter \& tracker & Basic account info                & 6                 & 486                       & 2                     \\
        YouTube on Xbox Live                        & YouTube                           & 6                 & 420                       & 2                     \\
        Fetch Rewards                               & Basic account info, Gmail         & 6                 & 159                       & 3                     \\
    \end{tabular}
\end{table*}

}

\newcommand{\tabSSOTable}[0]{
\begin{table*}[h]
    \caption{The top 25 authorized SSO ranked by the number of authorized accesses to participants' Google accounts, and the number of permissions each SSO requested.\label{tab:ssos}}
    \footnotesize \centering
    \begin{tabular}{r|l|c|c}
    
                                                            &                                               & \textbf{Number}  & \textbf{Number of}    \\
        \textbf{SSO}                                        & \textbf{Permission Categories}                & \textbf{Authorized}    & \textbf{Permissions}  \\
        \toprule
        Movies Anywhere                                     & Basic account info, Google Play               & 167               & 3                     \\
        Honey                                               & Basic account info                            & 80                & 2                     \\
        Amazon Alexa                                        & Basic account info, Gmail, Google Calendar    & 72                & 4                     \\
        Quora                                               & Basic account info, Google Contacts           & 71                & 4                     \\
        Adobe                                               & Basic account info                            & 63                & 2                     \\
        Reddit                                              & Basic account info                            & 61                & 3                     \\
        Microsoft apps \& services                          & Gmail, Google Calendar,                       & 58                & 8                     \\
                                                            & Google Contacts, Google Drive                 &                   &                       \\
        Pinterest                                           & Basic account info                            & 58                & 2                     \\
        Windows                                             & Basic account info, Gmail,                    & 53                & 8                     \\
                                                            & Google Calendar, Google Contacts              &                   &                       \\
        Glassdoor                                           & Additional access, Basic account info         & 52                & 3                     \\
        The New York Times                                  & Basic account info                            & 49                & 2                     \\
        Doordash                                            & Basic account info                            & 46                & 2                     \\
        Spotify                                             & Basic account info                            & 44                & 2                     \\
        macOS                                               & Basic account info, Gmail, Google Calendar,   & 43                & 6                     \\
                                                            & Google Contacts, Google Hangouts              &                   &                       \\
        Quizlet                                             & Basic account info                            & 43                & 2                     \\
        Dropbox                                             & Basic account info, Google Contacts           & 41                & 4                     \\
        SAMSUNG Account                                     & Additional access, Basic account info         & 38                & 5                     \\
        Zoom                                                & Basic account info, Google Calendar,          & 35                & 5                     \\
                                                            & Google Contacts                               &                   &                       \\
        Shop: package \& order tracker                      & Basic account info, Gmail                     & 32                & 3                     \\
        Nest                                                & Additional access, Basic account info         & 31                & 3                     \\
        paypal.com                                          & Basic account info                            & 31                & 2                     \\
        Adobe Acrobat Reader: PDF Viewer, Editor \& Creator & Basic account info, Google Drive              & 30                & 5                     \\        
        Best Buy App                                        & Basic account info                            & 30                & 2                     \\
        Shop                                                & Basic account info, Gmail                     & 30                & 3                     \\
        SoundCloud                                          & Basic account info                            & 29                & 2                     \\
    \end{tabular}
\end{table*}
}


\newcommand{\figHowOftenReviewAppSSO}[0]{
\begin{figure}[t]
\centering
\includegraphics[width=\columnwidth]{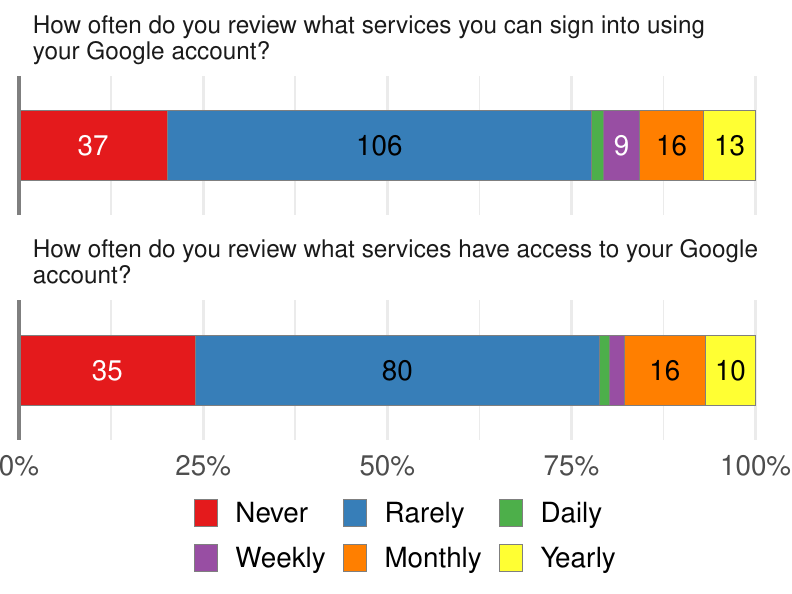}
\caption[How often do you review services bar plot]{\label{fig:howOftenReviewAppSSO-bar}
Over \SI{75}{\percent} of participants stated that they \emph{Never} or \emph{Rarely} review what services they can sign into using~(\ref{appendix:main-survey:Q2}) or have access to their Google account~(\ref{appendix:main-survey:Q4}).
}
\end{figure}
}

\newcommand{\figRemindAndReapprove}[0]{
\begin{figure}[t]
\centering
\includegraphics[width=\columnwidth]{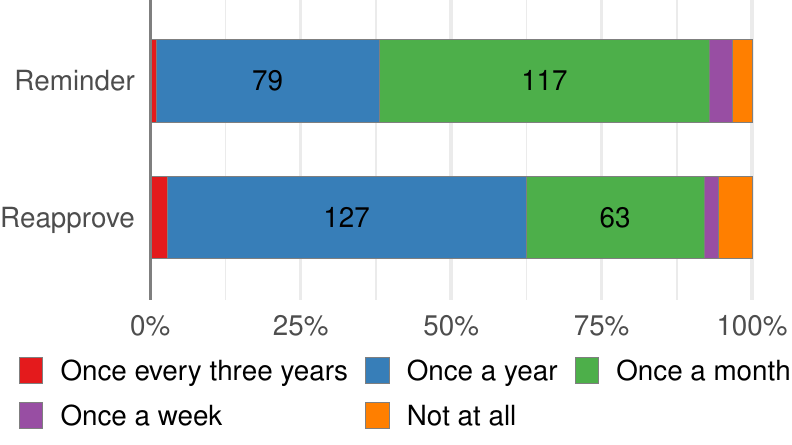}
\caption[Reminder to review, require reapproval bar plot]{\label{fig:remindAndReapprove-bar}
A majority of participants stated that they wanted an email reminder to review their \enquote{Apps with access to your account} page \emph{Once a month}~(\ref{appendix:main-survey:Q22}) and to reapprove apps \emph{Once a year}~(\ref{appendix:main-survey:Q23}).
}
\end{figure}
}

\newcommand{\figBenefitConcernChange}[0]{
\begin{figure}[t]
\centering
\includegraphics[width=\columnwidth]{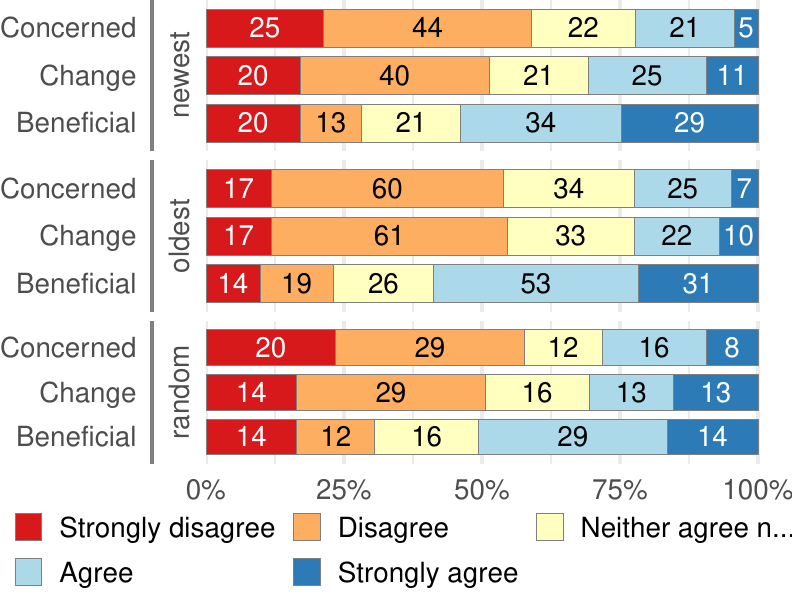}
\caption[Access benefits, concerns, and want to change bar plot]{\label{fig:benefitConcernChange-bar}
Full results of question (\ref{appendix:main-survey:Q10}). Most participants are not concerned about their third-party apps having access to their Google account. Over half of participants do not want to change the parts of their Google account that an app can access. A majority of participants agree that app access to their Google account is beneficial.
}
\end{figure}
}

\newcommand{\figUnderstandingLinkAccess}[0]{
\begin{figure}[t]
\centering
\includegraphics[width=\columnwidth]{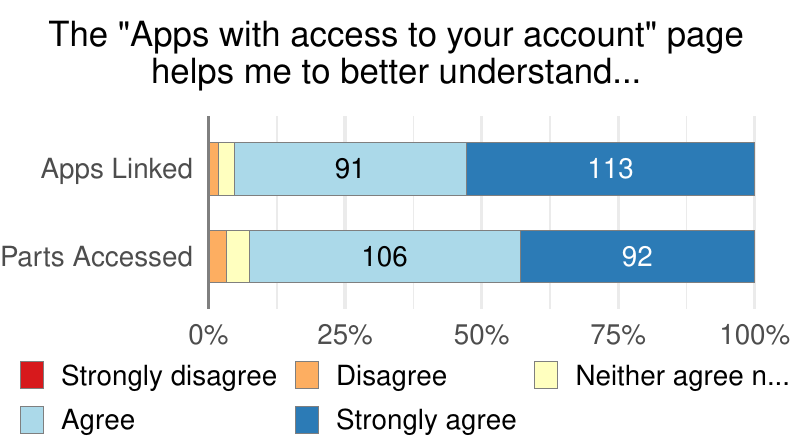}
\caption[Understanding linking and account access bar plot]{\label{fig:understandingLinkAccess-bar}
Over \SI{90}{\percent} of participants \emph{Strongly Agree} or \emph{Agree} that the enquote{Apps with access to your account} page helps them to better understand both which apps are linked to their account~(\ref{appendix:main-survey:Q15}), and what parts of their account apps can access~(\ref{appendix:main-survey:Q16}).
}
\end{figure}
}

\newcommand{\figChangeSettingsReviewApps}[0]{
\begin{figure}[t]
\centering
\includegraphics[width=\columnwidth]{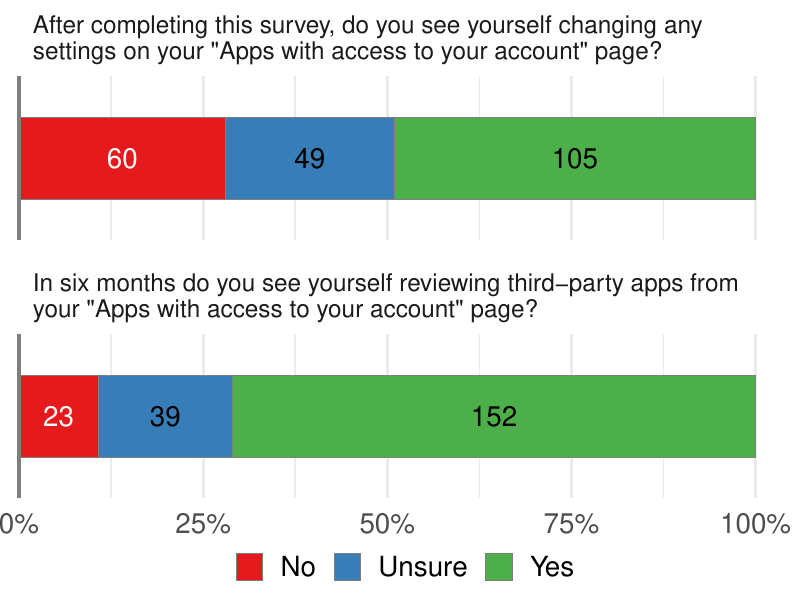}
\caption[Change settings and review apps bar plot]{\label{fig:changeSettingsReviewApps-bar}
Roughly \SI{50}{\percent} of participants stated that they would change their settings after the survey~(\ref{appendix:main-survey:Q17}), and a majority would review third-party apps in six months~(\ref{appendix:main-survey:Q18}).
}
\end{figure}
}

\newcommand{\figBeforeGrantingAccess}[0]{
\begin{figure}[t]
\centering
\includegraphics[width=\columnwidth]{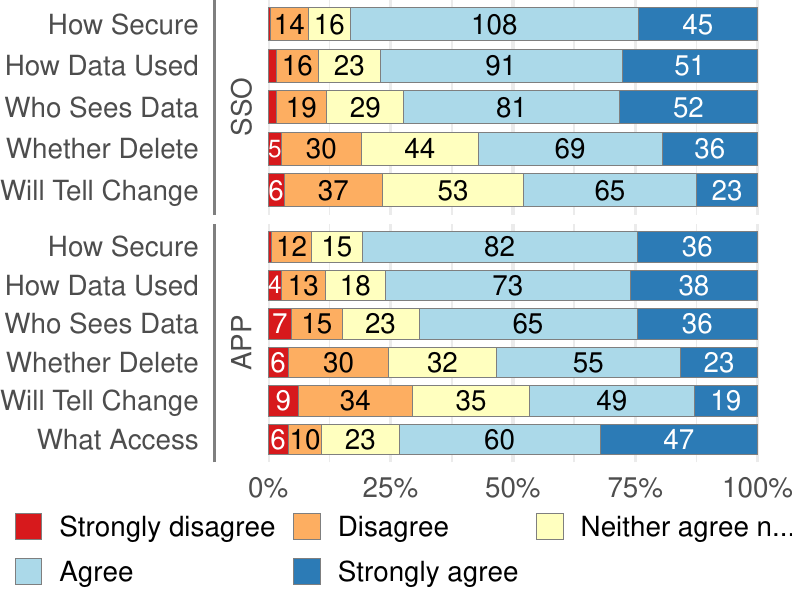}
\caption[Before granting access bar plot]{\label{fig:beforeGrantingAccess-bar}
Full results of questions (\ref{appendix:main-survey:Q1}) and (\ref{appendix:main-survey:Q3}) in which participants are asked what they consider before granting Google account access to a third-party APP or SSO.
}
\end{figure}
}

\newcommand{\figRecallSSOThirdParty}[0]{
\begin{figure}[t]
\centering
\includegraphics[width=\columnwidth]{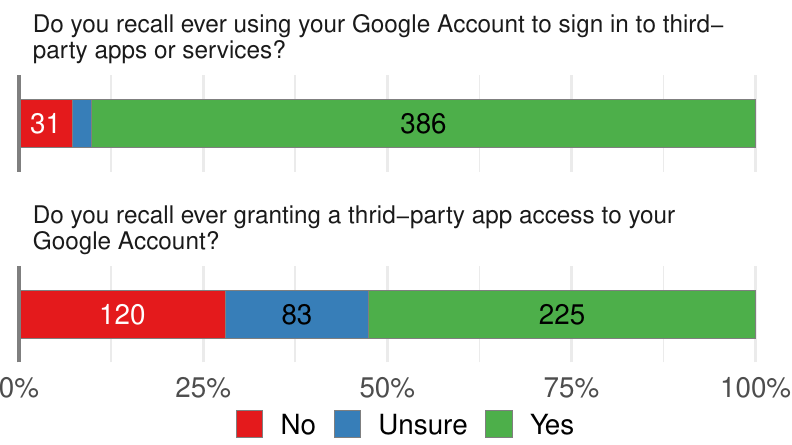}
\caption[Before granting access bar plot]{\label{fig:recallSSOThirdParty-bar}
Ninety percent of participants recall using their Google Account to sign in  (\ref{appendix:pre-survey:S6}) and over half recall granting a third-party app access to their Google Account (\ref{appendix:pre-survey:S10}).
}
\end{figure}
}

\newcommand{\figLastTimeUsingApp}[0]{
\begin{figure}[t]
\centering
\includegraphics[width=\columnwidth]{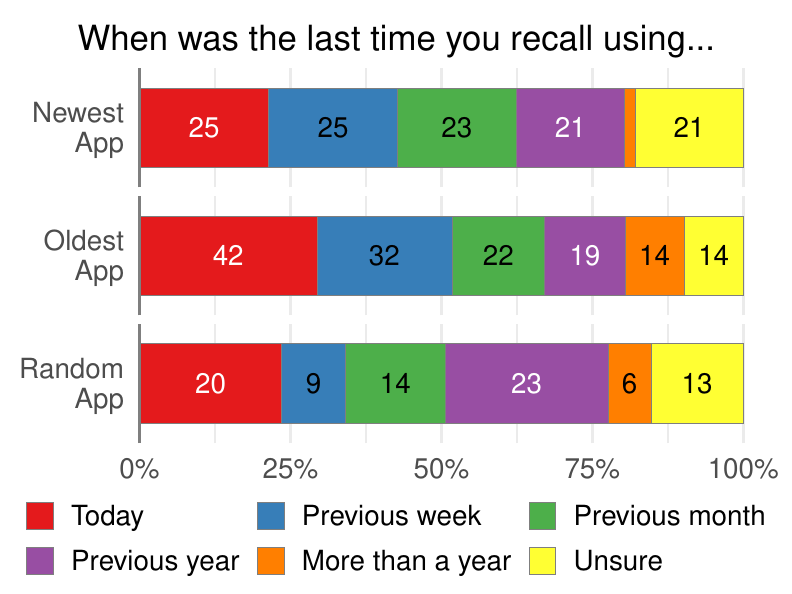}
\caption[Last time using app bar plot]{\label{fig:lastTimeUsingApp-bar}
Participants' oldest app was also their most recently used as over half report using their oldest app \emph{Today} or \emph{In the previous week} (\ref{appendix:main-survey:Q8}).
}
\end{figure}
}

\newcommand{\figAwareRecallKeep}[0]{
\begin{figure}[t]
\centering
\includegraphics[width=\columnwidth]{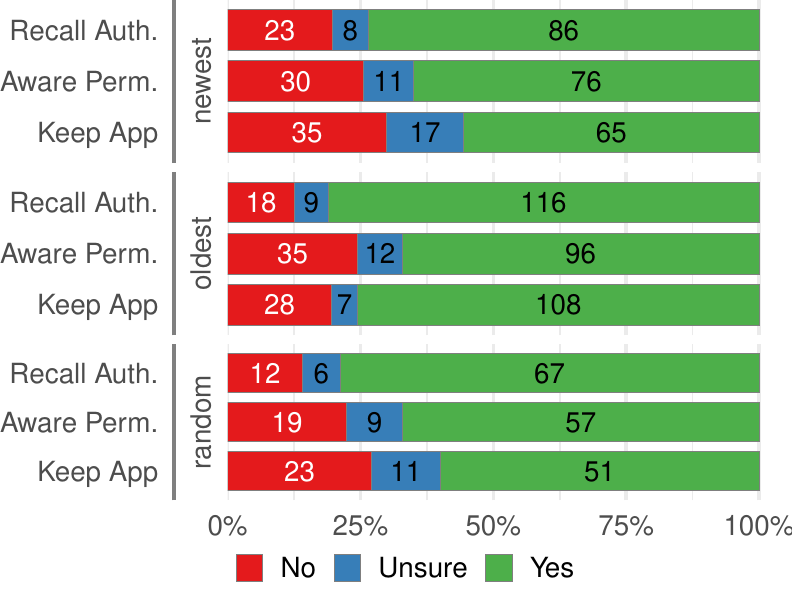}
\caption[Aware recall keep bar plot]{\label{fig:awareRecallKeep-bar}
Full results of questions (\ref{appendix:main-survey:Q5}), (\ref{appendix:main-survey:Q7}), and (\ref{appendix:main-survey:Q9}). Most participants recall authorizing their apps and are aware that their apps had permissions to access parts of their Google account. Their newest app is the mostly likely to be removed.
}
\end{figure}
}

\newcommand{\figApproveEveryBlockData}[0]{
\begin{figure}[t]
\centering
\includegraphics[width=\columnwidth]{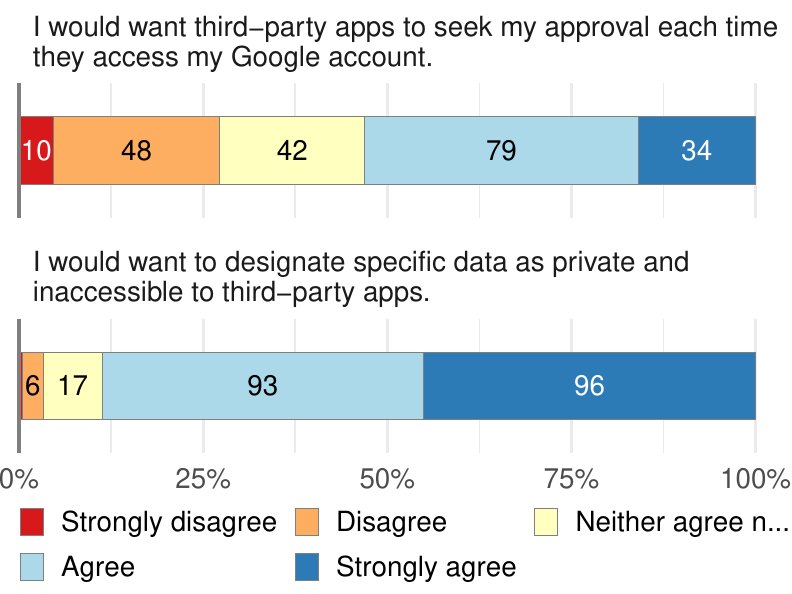}
\caption[Approve every block data bar plot]{\label{fig:approveEveryBlockData-bar}
Over half of participants \emph{Strongly Agree} or \emph{Agree} they want third-party apps to seek approval each time they access their Google account~(\ref{appendix:main-survey:Q25}), and 
roughly \SI{90}{\percent} of participants \emph{Strongly Agree} or \emph{Agree} they want to designate specific data as private and inaccessible to third-party apps~(\ref{appendix:main-survey:Q26}). \changes{One participant managed to submit a non-response, and is excluded from this data.}
}
\end{figure}
}

\newcommand{\figHowConfidentPermission}[0]{
\begin{figure}[t]
\centering
\includegraphics[width=\columnwidth]{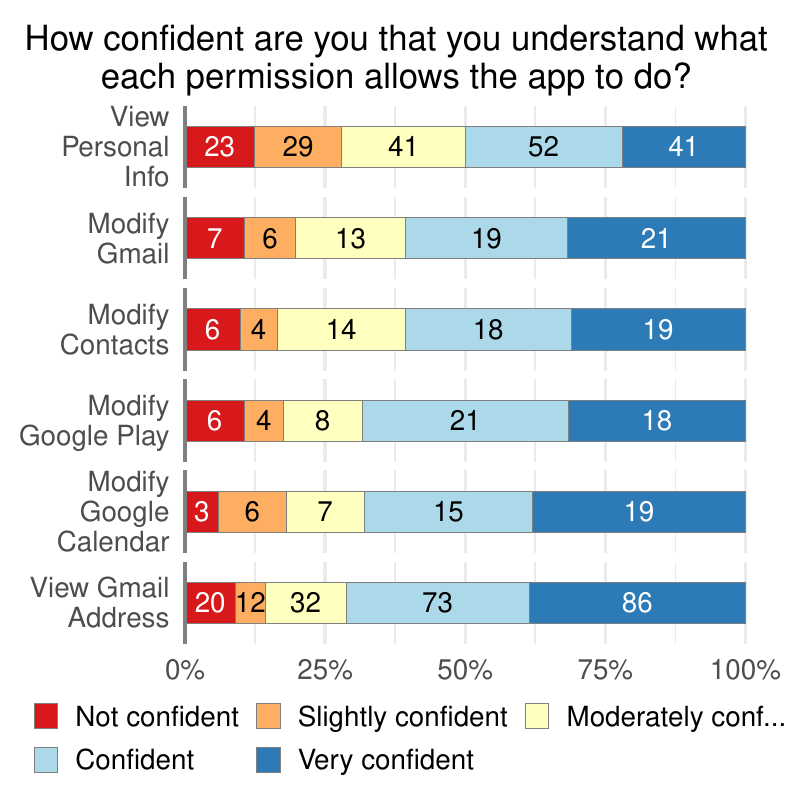}
\caption[How confident permission bar plot]{\label{fig:howConfidentPermission-bar}
Results for (\ref{appendix:main-survey:Q11}) for the top six most prevalent permissions, of which \enquote{View Personal Info} was the least understood and \enquote{View Gmail Address} was the most.
}
\end{figure}
}

\newcommand{\figHowNecessaryPermission}[0]{
\begin{figure}[t]
\centering
\includegraphics[width=\columnwidth]{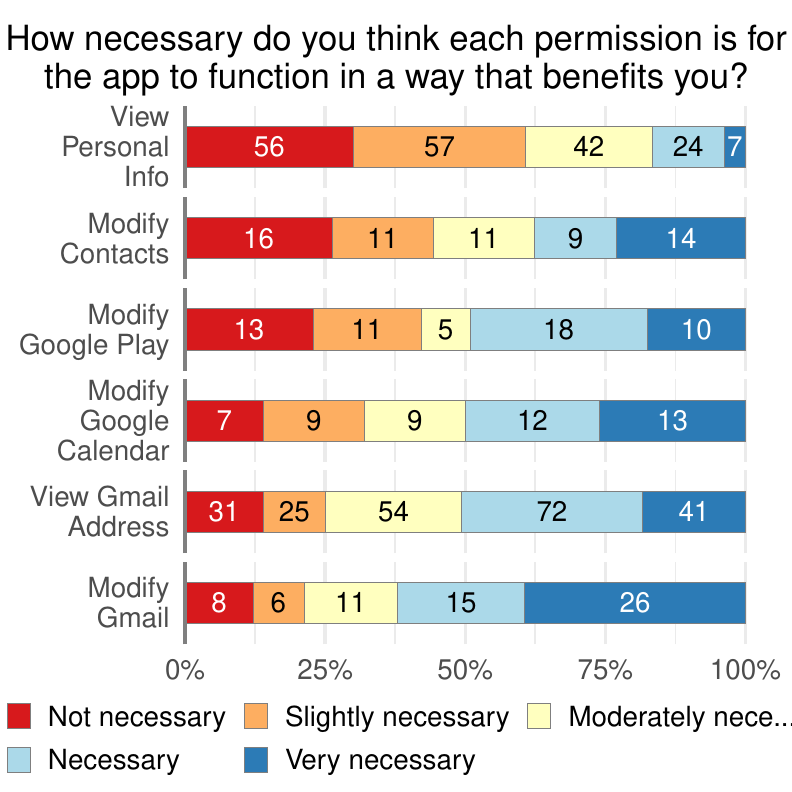}
\caption[How necessary permission bar plot]{\label{fig:howNecessaryPermission-bar}
Results for (\ref{appendix:main-survey:Q12}) for the top six most prevalent permissions, of which \enquote{View Personal Info} was considered the least necessary and \enquote{Modify Gmail} was the most.
}
\end{figure}
}

\newcommand{\figHowConcernedPermission}[0]{
\begin{figure}[t]
\centering
\includegraphics[width=\columnwidth]{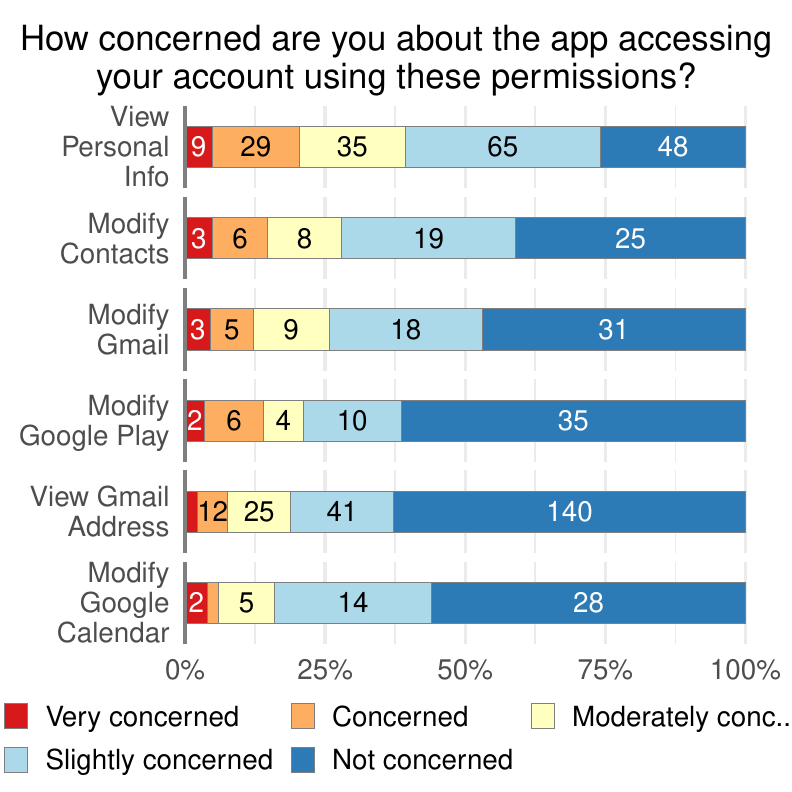}
\caption[How concerned permission category bar plot]{\label{fig:howConcernedPermission-bar}
Results for (\ref{appendix:main-survey:Q13}) for the top six most prevalent permissions, of which \enquote{View Personal Info} was the most concerning and \enquote{Modify Google Calendar} was the least.
}
\end{figure}
}
\begin{abstract}

\changes{Online services like Google} provide \changes{a variety of application programming interfaces (APIs). These online APIs enable authenticated third-party services and applications (apps)} to access a user's \changes{account data} for tasks such as single sign-on (SSO), calendar integration, \changes{and sending email on behalf of the user, among others}. Despite their prevalence, \changes{API access} could pose significant privacy and security \changes{risks,} where a third-party could \changes{have unexpected privileges to a} user's account. To gauge users' perceptions and concerns \changes{regarding} third-party apps that \changes{integrate with online APIs}, we performed a multi-part online survey \changes{of Google users}. First, \changes{we asked} $n=432$ participants to recall if and when they allowed third-party access to their Google account: 89\% recalled using at least one SSO and 52\% remembered at least one third-party app. In the second survey, we re-recruited $n=214$ participants to ask about specific \changes{apps and SSOs they've authorized on} their own Google \changes{accounts. We collected in-the-wild data about users' actual SSOs and authorized apps: 86\% used Google SSO on at least one service, and 67\% had at least one third-party app authorized. After examining their apps and SSOs,} participants expressed the most concern about \changes{access to} personal information \changes{like email addresses} and other publicly shared info. \changes{However, participants} were less concerned with broader---and perhaps more invasive---access to calendars, emails, or cloud storage (as needed by third-party \changes{apps}). This discrepancy may be due in part to trust transference to apps that integrate with Google, forming an implied partnership. Our results suggest \changes{opportunities} for design improvements to the current third-party management tools offered by Google; for example, tracking recent access, automatically revoking access due to app disuse, and providing permission controls.


\end{abstract}
\section{Introduction}\label{sec:introduction}


\changes{The 2018 Cambridge Analytica scandal~\cite{fb-user-data-privacy} prompted the scrutiny of third-party apps that integrate with online application programming interfaces (APIs). This came about when an online personality quiz used the Facebook API to collect detailed personal information from millions of unsuspecting quiz-takers and their friends. In response, Facebook restricted how third-parties can use its API~\cite{fb-restricting-data-access}. However, Facebook is not unique among online services that allow third-party apps to leverage user account data via APIs.}


\changes{Likewise, Google has APIs that allow third-party services to use existing user account data. For example, Google's single sign-on (SSO)~\cite{google-sso} lets users log onto participating third-party services with their already-existing Google credentials. Other major online services like Apple~\cite{apple-sso}, Twitter~\cite{twitter-sso}, and Facebook~\cite{facebook-sso} offer similar SSO capabilities using their accounts.}


\changes{The Google API also exposes functionality from various Google products. For instance, APIs let authenticated third-party apps to interact with Google Calendar entries or Gmail correspondence on behalf of the user. This particular integration is what enables iOS's built-in Calendar app to display and edit a user's Google Calendar events, among others.}


\changes{Despite many benefits, granting programmatic access to one's online account can pose security and privacy risks, as highlighted by Cambridge Analytica. In 2018, Google proposed granular permissions for API authorizations to give users more control and mitigate these risks~\cite{google-granular-permissions}. As of September 2021, however, this updated design does not appear to be widely adopted. Google API integrations in popular services like Dropbox and Zoom still use all-or-nothing consent flows, and a spot check of the most popular apps on the Google Workspace Marketplace\footnote{https://workspace.google.com/marketplace/category/popular-apps} shows those too lack fine-grained permissions.} 

In this paper we explore how users consider security and privacy in light of third-party API access to their Google accounts \changes{given the disclosure and control mechanisms currently available}.
%
%
\changes{First, we surveyed $n=432$ participants to recall the last times they used Google SSO or authorized a third-party app access to their Google account data. When recalling these, we also asked participants what factors they considered before granting access.} Of the 432 participants, the vast majority (89\%) recalled using SSO, but only half (52\%) recalled granting the third-party access to their Google account.   


\changes{We then invited $n=214$ participants from the first survey to return for a follow-up survey. As part of this second survey, participants installed a browser extension that parsed entries in their Google account's ``Apps with access to your account'' dashboard.\footnote{https://myaccount.google.com/permissions} Based on this data, we asked participants about specific apps they currently have installed on their Google account. From the browser extension, we observed
1,010 third-party services that use Google SSO and 455 third-party apps that integrate with APIs for various Google services. Of the observed third-party apps}, nearly half require two or more permissions accessing the participants' Google account. The most common permission is modifying Google Play Game activity (223 instances), followed by viewing primary Google email address (189), and viewing personal info (177). 


Participants were overall only {\em Slightly concerned} or {\em Not concerned} about the access granted to third-party apps, but showed the most concern about apps viewing personal info; 39\% were {\em Very concerned}, {\em Concerned}, or {\em Moderately concerned}. Interestingly, such information is perhaps less of a privacy and security risk than third-party apps that can modify/view contacts, email, calendar events, or cloud storage.  The \changes{relative lack of} concern \changes{with} these permissions could be attributed to a transference of trust to Google, as evidenced by open-ended responses where participants indicated that they believe Google is properly vetting these accesses. 


We surveyed participants about \changes{the specific apps on their accounts} and asked if they wished to keep or remove account access \changes{for those apps}. A logistic regression revealed that participants were $5.8\times$ more likely to want to remove access for an app when they wished to change which Google account data the app can access. Additionally, they were $5.9\times$ more likely to keep access when the app was recently used, and $6.0\times$ more likely to keep access when they viewed the app as beneficial. However, 79\% and 78\% of participants indicated that they currently {\em Rarely} or {\em Never} review their apps and SSOs, respectively. After viewing their third-party accesses \changes{as part of our survey} though, the vast majority (95\%) of participants indicated they would want reminders to review those at least {\em Once a year}.

These findings suggest a significant opportunity to improve how users interact with third parties with \changes{programmatic} access to their \changes{accounts} by helping users to identify and remove less frequently used apps/SSO in an automated way, or to simply \changes{revoke access after a period of disuse}. Similarly, Google could require regular re-approval of access, perhaps yearly so as not to be too disruptive. Additionally, many participants articulated a desire for controls of the permissions for third-party apps. This would allow users to better limit which aspects of their Google account each app/SSO can access with respect to the benefit being provided, rather than forcing them to accept an all-or-nothing approach.



%
\section{Background and Related Work}\label{sec:related-work}

Russell, et al.~\cite{russell-apis} characterize online APIs as \changes{among:} \textit{content-focused APIs} that provide data; \textit{feature
APIs} that integrate existing software functionality
from elsewhere; \textit{unofficial APIs} that (unintentionally) expose internal interfaces; and \textit{analytic APIs} that track user experiences. Here,  we focus on Google's content-focused and feature APIs that enable third-party developers to register apps with Google that can perform operations on behalf of a user.  Most services, including Google's, use the OAuth standard~\cite{oauth} to delegate and manage these authorizations. OAuth has been the focus of much security research~\cite{bai2013authscan,sun2012devil,zuo2017authscope}, and in this paper we do not investigate the security of OAuth directly but rather user awareness and concerns for such delegations.

While we primarily focus on third-party apps, we also consider SSO services as a form of third-party apps with limited functionality. Bauer et al.~\cite{bauer2013comparison} looked at willingness to use the SSOs of Google, Facebook and other services, finding that there were concerns with information sharing through SSO, despite messaging. We find similar concerns in our study.  Ghasemisharif 
et al.~\cite{ghasemisharif2018single} studied SSO with respect to potential for account hijacking. The authors also measured the prevalence of SSOs, finding that Facebook is the most prevalent SSO service, followed by Twitter and Google.
Hu et al.~\cite{hu2014application} investigated SSOs in the context of online social networks and how apps can complete an impersonation attack. And Zhou et al. completed automated vulnerability testing of SSO on the web~\cite{zhou2014ssoscan}. Here we assume that the SSO is properly implemented and instead focus on user perceptions of sharing information with third-parties via SSO services.


Prior work on third-party apps have mostly focused on the Facebook ecosystem. Felt et al.~\cite{adrienne2008privacy}  examined 150
Facebook platform apps in 2008, finding that 90\% of the examined apps have unnecessary access to private data.
%
%
Huber et al.~\cite{huber-appinspect}
developed a method to analyze privacy leaks in Facebook apps at scale by leveraging client-side \texttt{iframes} to capture network traffic. 
Google third-party apps do not necessarily operate client-side.
%
%
More recently, Farooqi et al.~\cite{farooqi2020canarytrap} used ``honeytoken'' email addresses (\ie, auto-generated
accounts on an email server that the researchers control) to detect
Facebook apps inappropriately collecting and using those addresses. Such a method could also be used for Google third-party apps but was not the primary focus of this research. 


Our work is also related to prior research on permission management \changes{for online APIs}. Similar to Wang et al.~\cite{Wang2011}, who analyzed the permissions requested by Facebook API apps at install time, we explore the permissions requested by third-party apps that integrate with Google's API. Prior work explored a subset of these permissions on Google~\cite{reyesapi}. A lack of centralization for third-party apps means there is far from comprehensive coverage. Our work expands on this effort with in-the-wild observations of apps authorized on actual users' Google accounts.


Permissions have been extensively studied in the context of smartphone apps. Notably, Felt et al.~\cite{felt2011android, felt2012b} examined
Android apps and found that one-third are overprivileged, and Wijesekera et al.~\cite{wijesekera-remystified}
surveyed Android users' perceptions of app permissions, 80\% of whom  wished to deny at least one permission. 
With the shift towards runtime, ask-on-first-use permission requests~\cite{android-request,ios-request}, Wijesekera et al. in 2017~\cite{wikesekera2017feasibility} developed a classifier to predict user permission preferences by taking into account  the context of the permission request. Likewise, Smullen, et al. in 2020~\cite{smullen2020best}
built a classifier that demonstrated users are sensitive to the purpose of particular permissions requested by apps.
Our results show that similar contextualization has impact on users' perceptions of permissions, and recommend moving towards auto-review and auto-revocation models for third-party apps as a whole and for individual permissions. 

Finally, this research is also related to work on privacy and transparency dashboards for online services. These have been both  extensively proposed and explored in the literature~\cite{janic-13-tets-overview,raschke-17-usable-pri-dashboard,murmann-17-survey-tets,zimmermann-14-privacy-dashboards,rao-14-what-do-they-know,schnorf-14-transparency-trust,tschantz-18-accu-google-ad,weinshel-19-places-you-been,herder-20-privacy-dashboard,earp-16-my-activity,farke-21-privacy-dashboards}. 
The ``Apps with access to your account'' page, \changes{to which we direct participants}
in the survey, functions similarly to other transparency dashboards; however, this dashboard offers less functionality than other dashboards. We explore ways to improve management of third-party apps through the web interface in Section~\ref{sec:discussion}.

\section{Method}\label{sec:method}


We begin by describing the two surveys. The first survey asked participants to recall prior experiences with third-party apps and SSOs. 
The second survey \changes{leveraged} a custom browser extension and asked participants to respond to the specific SSOs and third-party apps \changes{currently authorized on} their Google account. In the remainder of this section, we \changes{detail our study procedures}, describe \changes{how we recruited participants}, \changes{discuss} ethical considerations \changes{of our study}, and outline the limitations of our approach.

\paragraph{First Survey.}
Below, the first survey is described. The full survey can be found in \autoref{appendix:pre-survey}.

%
\begin{enumerate}[leftmargin=*,noitemsep]
\item Informed Consent: Participants consented to the study.
\item Google Account Use: Participants were asked if they have a Gmail account (as surrogate for a Google account), if it is their primary Google account with sole ownership, and the age of the account. 
Questions:~\ref{appendix:pre-survey:S1}--\ref{appendix:pre-survey:S4}. 
\item Familiarity with ``Sign in with Google'': Participants were provided with contextual information from Google's documentation~\cite{google-sso} alongside a screenshot of a ``Sign in with Google'' button (taken from Yelp) and asked about their recent experiences using their Google account to sign into a third-party app or service.
Questions:~\ref{appendix:pre-survey:S6}--\ref{appendix:pre-survey:S9}. 
\item Familiarity with Third-Party App Account Access: Participants were provided with contextual information from Google's documentation~\cite{google-tp-apps} alongside a screenshot of a third-party app's Google API authorization screen (taken from a generic app). Next, participants were asked about their recent experiences granting a third-party app access to their Google account.
Questions:~\ref{appendix:pre-survey:S10}--\ref{appendix:pre-survey:S13}. 
\item IUIPC-8: Participants answered the Internet users' information privacy concerns (IUIPC-8) questionnaire~\cite{gross2021validity}, to gain insights into participants' privacy concerns.
\item Demographics: Participants were asked to provide demographic information, such as age, identified gender, education, and technical background.
Questions:~\ref{appendix:pre-survey:D1}--\ref{appendix:pre-survey:D4}. 
\end{enumerate}

\paragraph{Second Survey.} 
The second survey recruited from those who completed the first survey. We used the following inclusion criteria \changes{to ensure participants have active Google accounts with SSOs and/or third-party apps}:
\begin{enumerate*}[label=(\roman*)]
\item the participant has a Gmail account,
\item the participant uses the Gmail account as their primary Google account,
\item the participant has sole ownership of their Google account, 
\item the participant has used their Google account for more than one month,
\item the participant correctly answered the attention checks.
\end{enumerate*}

%
Below, we describe each part of the second survey in detail, and the full survey can be found in \autoref{appendix:main-survey}.
\begin{enumerate}[leftmargin=*,noitemsep]
\item Informed Consent: Participants consented to the main study, which included notice that they would be asked to install a web browser extension that would access their Google's \enquote{Apps with access to your account} page.
\item Extension Installation: Participants installed the browser extension that locally parses third-party apps and SSOs and displays specific apps to the participant as part of the survey. 
The extension also recorded aggregate information about the number of SSO and API authorizations and the date of the oldest and newest authorization.
\item Explore Apps with Access to Your Account: We provided participants with a brief descriptive introduction  and then directed them to explore their Google \enquote{Apps with access to your account} page for one minute. This interaction was managed by the browser extension with an overlay banner and restricted navigation away from the page. 
%
\item Account Access Questions: Participants were asked what they consider 
before allowing third parties access to their Google account and services and how often that access is reviewed.
%
%
Questions:~\ref{appendix:main-survey:Q1}--\ref{appendix:main-survey:Q4}. 
\item Keep or Remove: Each participants was shown all their Google account authorizations and if they wished to keep or remove the authorization.
%
%
Question:~\ref{appendix:main-survey:Q5}
\item App Permissions: Participants were asked about the permissions for their newest, oldest, and a random third-party app \changes{to investigate potential impacts of installation time on concern, benefit, and recall.}
%
Questions:~\ref{appendix:main-survey:Q7}--\ref{appendix:main-survey:Q14}. 
\item Reflections: Participants were asked to reflect on their understanding of the Google \enquote{Apps with access to your account} page and if they would change their behavior as a result of that interaction. 
%
Questions \ref{appendix:main-survey:Q15}--~\ref{appendix:main-survey:Q20}.
%
%
%
\item Feature Improvements: Participants provided suggestions for improving  Google's \enquote{Apps with access to your account} page.
%
Questions \ref{appendix:main-survey:Q21}--\ref{appendix:main-survey:Q26}.
\item Uninstall Extension: Upon completing the survey,  participants were instructed to remove the browser extension.
\end{enumerate}

\paragraph{Recruitment and Demographics.}

We recruited \num{432}~participants via \emph{Prolific}\footnote{\url{https://www.prolific.co} - Prolific participant recruitment service, as of \today.} for the first survey between March 31, 2021 and April 20, 2021. 
After applying our inclusion criteria, \num{399}~participants qualified for the second study, and \num{214} returned to complete the second survey.
Participants received \SI{1.00}[\dollar]{\USD} and \SI{3.00}[\dollar]{\USD} for completing the first and second survey, respectively, and it took, on average, \num{8}~minutes and \num{13}~minutes to complete, respectively.

Thirty-one percent of first survey participants were between \numrange[range-phrase = --]{18}{24} years old, \SI{36}{\percent} were between \numrange[range-phrase = --]{25}{34} years old, and \SI{33}{\percent} were \num{35} years or older.
The identified gender distribution for the first survey was \SI{50}{\percent} men, \SI{47}{\percent} women, and \SI{3}{\percent} non-binary, self-described, or choose not disclose gender.

Thirty-one percent of second survey participants were between \numrange[range-phrase = --]{18}{24} years old, \SI{37}{\percent} were between \numrange[range-phrase = --]{25}{34} years old, \SI{31}{\percent} were \num{35} years or older, \changes{and \SI{1}{\percent} chose not to disclose their age}.
The identified gender distribution for the second survey was \SI{50}{\percent} men, \SI{46}{\percent} women, and \SI{4}{\percent} non-binary, self-described, or choose not disclose gender.
Participant characteristics are presented in Table~\ref{tab:characteristics}. 

Additional demographic information can be found in Appendix~\ref{appendix:additional}.

\begin{table}[t]
\centering
\small
\caption[Characteristics of the study participants]{\label{tab:characteristics}
Demographic and IUIPC data collected at the end of the first survey.
}
\renewcommand{\arraystretch}{0.6}
\begin{tabular*}{\columnwidth}{@{}>{\bfseries}ll@{\extracolsep{\fill}}*{4}{r}}
\toprule
&                & \multicolumn{2}{c}{\textbf{First}} & \multicolumn{2}{c}{\textbf{Second}} \\
&                & \multicolumn{2}{c}{\textbf{Survey}} & \multicolumn{2}{c}{\textbf{Survey}} \\
&                & \multicolumn{2}{c}{\emph{(n = \num{432})}} & \multicolumn{2}{c}{\emph{(n = \num{214})}} \\
\midrule
\multirow{6}{*}{\rotatebox[origin=c]{90}{Gender}}
& & \multicolumn{1}{c}{\textbf{n}} & \multicolumn{1}{c}{\textbf{\%}} & \multicolumn{1}{c}{\textbf{n}} & \multicolumn{1}{c}{\textbf{\%}} \\
\cmidrule{3-6}
%

%
& Woman          & 204    &  47     &   99    &  46 \\
& Man            & 216    &  50     &  107    &  50 \\
& Non-binary     &  11    &   3     &    7    &   3 \\
& No answer      &   1    &   0     &    1    &   1 \\
\midrule
\multirow{6.5}{*}{\rotatebox[origin=c]{90}{Age}}
%
%
%
& 18--24         & 132    & 31  &  66    & 31  \\
& 25--34         & 157    & 36  &  80    & 37  \\
& 35--44         &  75    & 17  &  32    & 15  \\
& 45--54         &  44    & 10  &  20    &  9  \\
& 55+            &  23    &  6  &  15    &  7  \\
& No answer      &   1    &  0  &   1    &  1  \\
\midrule\multirow{7}{*}{\rotatebox[origin=c]{90}{IUIPC}}
& & \multicolumn{1}{c}{\textbf{Avg.}} & \multicolumn{1}{c}{\textbf{SD}} & \multicolumn{1}{c}{\textbf{Avg.}} & \multicolumn{1}{c}{\textbf{SD}} \\
\cmidrule{3-6}
%
%

& Control        &   5.9 &  1.0 &   5.8 &  1.0 \\
& Awareness      &   6.6 &  0.7 &   6.6 &  0.8 \\
& Collection     &   5.4 &  1.2 &   5.3 &  1.2 \\
\cmidrule{2-6}
& IUIPC Combined &   5.8 &  0.8 &   5.8 &  0.8 \\
\bottomrule
\end{tabular*}
\end{table}

\paragraph{Analysis methods.}
When performing quantitative analysis, descriptive or statistical tests, the analysis is provided in context. For qualitative responses, we use open coding to analyze responses to open-text questions. First, a primary coder from the research team crafted a codebook and identified descriptive themes for each question. A secondary coder coded a \SI{20}{\percent} sub-sample as a consistency check,  providing feedback on the codebook and iterating with the primary coder until inter-coder agreement was reached (Cohen's $\kappa > 0.7$\changes{, $mean = 0.81$, $sd = 0.05$}). 

\paragraph{Ethical Considerations.}
%
The study protocol was approved by The George Washington University Institutional Review Board (IRB) with approval number \texttt{NCR202914}, and throughout the process we considered the sensitivity of participants' Google app authorization data at every step.
All aspects of the survey requiring access to the actual Google \enquote{Apps with access to your account} page was administered \emph{locally} on the participant's machine using the browser extension.
All participants were informed about the nature of the study prior to participating and consented to participating in both surveys. At no time did the extension or the researchers have access the participants' Google password or to any other Google account data, and all collected data is associated with random identifiers.
%

\paragraph{Limitations.}

As an online survey, we are limited in that we cannot probe deeply with follow-up questions to understand the full range of responses. We attempt to compensate for this limitation by performing thematic coding across many responses to capture general opinions and feelings when interacting with third-party apps and SSOs that have access to their Google account.

Additionally, we are limited in our recruitment sample, which is generally younger than the population as a whole. Yet, we argue that despite this limitation, our results offer insights into user awareness of third-party apps and SSO access to their Google account, as well as other online services with third-party APIs. \changes{We note that prior work by Redmiles, et al.~\cite{redmiles-19-generalize} suggests online studies about privacy and security behavior can still approximate behaviors of populations.}

Some results may be affected by social desirability bias, where participants might indicate behavior that they believe we (the researchers) expect them to embody. For example, this may lead to participants over describing their awareness or recall of granting access to third-party apps or SSOs, or their intention to remove access. In these cases, one may view these results as a potential upper bound on true behavior.

\changes{Finally, we acknowledge our study only considers Google, even though many online services offer APIs. Still, we believe our findings are broadly applicable because of the consistency of the underlying mechanisms (i.e. OAuth scopes) and user consent flows (i.e., users grant install-time permissions with a dashboard for review) across many major providers.}

\section{Results}\label{sec:results}

In this section we present the results of the two surveys. We first describe the observed apps, SSOs, and their permissions for participants completing the second survey. Next, we explore participants' awareness and understanding of third-party apps and their permissions. We then offer results on participants' motivations to install or not install a third-party app and their intentions to change settings. Finally, we report on participants' reflections on their third-party apps and desired features for improving transparency and control over API access to their Google account.

\subsection{Measurements}
    
    




    \paragraph{Third-party app and SSO findings.}
    In the first survey, 432 participants self reported if  they recalled using Google's SSO (\ref{appendix:pre-survey:S6}) or granting third-party apps (\ref{appendix:pre-survey:S10}) access to their Google account. We found that 89\% ($n = \num{386}$) of participants recalled using SSO to sign into a third-party service.
    Furthermore, 52\% ($n = \num{225}$) recalled granting a third-party app access to their Google accounts. 
    %
    \figRecallSSOThirdParty
    See \autoref{fig:recallSSOThirdParty-bar}. 
    

    Additionally, during the first survey, we asked participants to recall the latest app or service they signed into using their Google account (\ref{appendix:pre-survey:S7}), and classified their responses into various categories based on the app or service. 
    We searched repositories of available third-party apps, e.g., Google Play or Google Workspace Marketplace, to determine the proper app category. 
    The top categories include shopping ($n = \num{51}$; 12\%), social media ($n = \num{42}$; 10\%), gaming ($n = \num{38}$; 9\%), food ($n = \num{34}$; 8\%) and entertainment ($n = \num{28}$; 6\%).

    Among the 214 second survey participants, a majority ($n = 184$; 86\%) have at least one SSO linked to their Google account and 67\% ($n = 143$) have at least one third-party app with Google account access. 
    Via the custom browser extension installed by the participants, we observed a total of 1,010 unique SSOs and 455 unique third-party apps accessing participants' Google accounts. For those who have at least one SSO ($n = 184$), the average number of SSOs per participant is 13 (median = 9.5; sd = 12).
    In comparison, those who have at least one third-party app ($n = 143$) have an average number of six third-party apps per participant (median = 3; sd = 6.7).
    %
    
    Third-party apps were authorized to access participants' Google accounts for an average of 285 days (median = 142; sd = 375). The maximum number of days authorized was 2,519 days and the minimum was one day. 
    The highest number of SSOs linked to a single participant's Google account is 65,
    and one participant had 49 third-party apps, the most observed.
    For a list of study participants' most installed apps please refer to  \autoref{tab:apps} in \autoref{appendix:additional}.


    %
    \paragraph{Associated account access permissions.}
    Among the 1,010 distinct SSOs, we recorded 114 unique associated permissions.
    Moreover, we cataloged 144 unique permissions requested across the 455 distinct third-party apps.
    The average number of permissions per SSO was three (median = 2; sd = 1.5);
    per third-party app, the average number of permissions was three (median = 2; sd = 2.2).
    
    The third-party app with the greatest number of permissions was \emph{Health Sync} with 19 permissions.
    \emph{Health Sync} was followed by: \emph{autoCrat} (14), \emph{FitToFit} (14), \emph{DocuSign GSuite Add-on} (13), \emph{Zero - Simple Fasting Tracker} (13), and \emph{Yahoo!} (12).
    Only a few ( 1\%; $n = 12$) SSOs ask for a single permission,  
    while 36\% ($n = 166$) of third-party apps have only one permission.
    %
    %
    Additionally, 78\% ($n = 792$) of SSOs have two or fewer permissions while 56\% ($n = 255$) of third-party apps have two or fewer permissions.

    The most common observed permission was \enquote{Create, edit, and delete your Google Play Games activity} ($n = 223$). 
    This was followed by: \enquote{See your primary Google Account email address} ($n = 189$), \enquote{See your personal info, including any personal info you've made publicly available} ($n = 177$), \enquote{See, create, and delete its own configuration data in your Google Drive} ($n = 71$), and \enquote{Associate you with your personal info on Google} ($n = 44$). 
    See \autoref{tab:app-permissions} for a list of study participants' most authorized third-party account access permissions.

\subsection{Awareness and Understanding}
\label{awarenessUnderstandingSubsection}

\figAwareRecallKeep
%
%

\paragraph{Awareness of Third-party Apps and SSOs.}
%


\changes{In the second survey, we used the browser extension to show participants their} newest, oldest, and a random third-party app. \changes{For participants with only two apps, we considered those apps as their oldest and newest. And for participants with only one app, we considered it as their oldest. This results in imbalanced participants with oldest apps $n_{old}=143$, newest apps $n_{new}=117$, and random apps $n_{rand}=85$.}

Participants were first asked if they recalled authorizing each app (\ref{appendix:main-survey:Q7}).
%
%
%
%
\changes{Among participants with at least one app}, \SI{33}{\percent} ($n = 47$) could not recall authorizing at least one of them.
The oldest installed app was recalled the most often \SI{81}{\percent} ($n = \num{116}$), followed by the randomly selected app at \SI{79}{\percent} ($n = \num{67}$), and finally the newest app at \SI{74}{\percent} ($n = \num{86}$).
%
%
There were no significant differences between the apps shown with respect to positive recall compared to negative or unsure responses ($\chi^2=0.27,p=0.87$). (See \autoref{fig:lastTimeUsingApp-bar} displays the full details.)

When asked when they last used these apps (\ref{appendix:main-survey:Q8});
over half (51\%; $n = 74$) of the participants reported using their oldest app \emph{Today} or \emph{In the previous week}. Whereas 43\% ($n = 50$) report using their newest app, and only 34\% ($n = 29$) report using their random app, over that same time period. There were no statistical differences (using a Kruskal-Wallace test $H=2.15,p=0.34$) between reported last use of apps when comparing newest, oldest, and random apps. 

Subsequently, we asked participants if prior to seeing the details about their newest, oldest, and a randomly chosen app in the survey, they were aware that the app had permissions to access their Google account data (\ref{appendix:main-survey:Q9}).
%
%
%
%
%
Forty-nine percent ($n = 70$) of participants with third-party apps were not aware, for at least one of those apps, that the app had permissions to access parts of their Google account data.
%
%
Participants were more likely to be unaware of the Google account access permissions of their newest app (35\%; $n = \num{41}$), followed by their oldest app (33\%; $n = \num{47}$), and random app (33\%; $n = \num{28}$).
Again, though, there were no significant differences between awareness of the apps ($\chi^2=0.032,p=0.98$).
\autoref{fig:awareRecallKeep-bar} shows the full results of app recall and awareness.

In the first survey, when participants were asked if they recalled using Google's SSO (\ref{appendix:pre-survey:S6}), 10\% ($n = 42$) responded \emph{No} or \emph{Unsure}. Yet, 16 of the 19 (84\%) of those same participants who completed the second survey actually did have a SSO linked with their Google account. We also asked whether they recalled granting a third party access to their Google account (\ref{appendix:pre-survey:S10}), and 47\% ($n = 203$) answered \emph{No} or \emph{Unsure}. However, 52 of 95 (55\%) of those very same participants who completed the second survey in fact had granted a third-party app access to their account.

\figLastTimeUsingApp
\figBenefitConcernChange
\paragraph{Benefits From and Concerns For Third-Party Apps.} 
%
%
Participants selected on a 5-point Likert agreement scale to indicate the benefits of their newest, oldest and randomly picked apps (\ref{appendix:main-survey:Q10}). 
%
%
Eighty-two percent ($n = 117$) of participants \changes{with apps} \emph{Agree} or \emph{Strongly agree} that at least one of their third-party apps is beneficial.
The oldest app was reported as the most beneficial with 59\% ($n = \num{84}$) who \emph{Agree} or \emph{Strongly agree}, followed by 54\% ($n = \num{63}$) for the newest app, and 51\% ($n = \num{43}$) for the random app. However, there were no significant differences across responses based on a Kruskal-Wallace test ($H=2.12,p=0.34$). 
Additionally, participants were asked whether they were concerned about their newest, oldest and random apps having access to their Google accounts (\ref{appendix:main-survey:Q10}). 
%
%
Fifty percent ($n = 71$) of participants \changes{with apps} \emph{Agree} or \emph{Strongly agree} that they were concerned with at least one of their third-party apps that can access their Google account.
Of the participants who were concerned, 28\% ($n = \num{24}$) \emph{Agree} or \emph{Strongly agree} to being concerned with their random app, 22\% ($n = \num{26}$) with their newest, and 22\% ($n = \num{32}$) with their oldest.
The full results for app benefit and concern are shown in \autoref{fig:benefitConcernChange-bar}, and again, there were no statistically significant differences ($H=1.58,p=0.45$).


\figHowConfidentPermission

\paragraph{Understanding App Access Permissions.}
We asked participants \changes{who have apps} to rate their confidence in understanding \changes{the permissions held by their third-party apps} (\ref{appendix:main-survey:Q11}).
\changes{Note that each participant had their own set of apps and permissions, so there is an imbalance in the number of participants surveyed for a given permission. Thus, we present permission-specific results as percentages, with full counts in the figures.}
Thirty-one percent ($n = 45$) of participants had at least one permission that they were \emph{Not confident} that they understood.
For the six most prevalent permissions requested by apps in our study, participants were \emph{Confident} or \emph{Very confident} (over \SI{50}{\percent}) in understanding each of them.
%
Participants were the most confident in understanding the permission \enquote{See your primary Google Account email address,} with \SI{33}{\percent} \emph{Confident} and \SI{39}{\percent} \emph{Very confident}.
This was also the most common permission surveyed, with 223 occurrences in third-party apps.
%
Conversely, participants were least confident in their understanding of the permission \enquote{See your personal info, including any personal info you've made publicly available,} with \SI{12}{\percent} \emph{Not confident}, \SI{16}{\percent} \emph{Slightly confident}.
There is also evidence in the qualitative data where a participants note that this permission is confusing because it does not sufficiently detail  what information is included in \enquote{personal info.}
This was the second most common permission surveyed, with 186 occurrences in third-party apps.
Results for the top six most prevalent permissions can be found in \autoref{fig:howConfidentPermission-bar}. Statistical comparisons were not performed due to the imbalance between groups for which permissions were surveyed.

\figHowNecessaryPermission

\paragraph{Necessity of Access Permissions.} 
We asked participants to report the necessity of each permission on a 5-point Likert scale for their newest, oldest and randomly selected third party app (\ref{appendix:main-survey:Q12}).
Sixty-one percent ($n = 87$) of participants had at least one permission that they reported was \emph{Not necessary}.
%

%
Among the six most prevalent permissions, participants found the permission \enquote{See your personal info, including any personal info you've made publicly available} to be the most unnecessary, with 30\% who stated that it is \emph{Not necessary} and 31\% who said it is only \emph{Slightly necessary}.
%
This is followed by the permission \enquote{See, edit, download, and permanently delete your contacts} in which 26\% said it was \emph{Not necessary} and 18\% said it is only \emph{Slightly necessary}.
\changes{At least 50\% of participants found the remaining prevalent permissions either \emph{Necessary} or \emph{Very necessary}.}
Participants rated functional permissions, such as \enquote{Read, compose, send, and permanently delete all your email from Gmail,} to be more necessary for the app to benefit them than data access permissions, like \enquote{See your personal info, including any personal info you've made publicly available.}
The top six most prevalent permission results for necessity can be found in \autoref{fig:howNecessaryPermission-bar}, and again, statistical comparisons were not performed due to the imbalance between groups for which permissions were surveyed. 

\figHowConcernedPermission

\paragraph{Concern for Access Permissions.} 
%
In \ref{appendix:main-survey:Q13} we asked the level of concern participants have about third-party apps accessing their account through various permissions.
Forty-six percent ($n = 66$) of participants had at least one permission that they were either \emph{Concerned} or \emph{Very concerned} about.
For five of the six most prevalent permissions requested by third-party apps observed in our study, over 70\% of participants answered \emph{Not concerned} or only \emph{Slightly concerned} about \changes{apps on their account having these permissions}.
%
\changes{The permission
\enquote{See your personal info, including any personal info you've made publicly available} had the highest concern}, with \SI{16}{\percent} \emph{Concerned}, \SI{5}{\percent} \emph{Very concerned}.
(See \autoref{fig:howConcernedPermission-bar}.)

\paragraph{Reasons for Concern.}
Participants also provided free responses describing any concerns they have with a third-party app, either the newest, oldest, or randomly chosen, having these access permissions (\ref{appendix:main-survey:Q14}).
Qualitative coding of the responses revealed that some participants expressed concern with the permissions held by their apps (newest, $n = \num{46}$; oldest, $n = \num{90}$; random, $n = \num{33}$).
The most common reasons for concern were access to personal or sensitive information, unnecessary access, ability to delete, and access to contacts and email. 
For example, P53 shared, \enquote{I don't want them having access to my personal information} (newest; YouTube on Xbox Live).
Examples of concern regarding unnecessary account access include:  \enquote{I'm a bit concerned about their access to my Youtube Channel, since it seems a bit unnecessary and excessive} (P170; newest; PlayStation Network) and \enquote{Not sure why it needs Google Drive access, it shouldn't be creating anything in there} (P93; random; Idle Island: Build and Survive).
Permissions that include the ability to delete files was often concerning, for instance, \enquote{I don't know if I want them to be able to delete stuff from my Google Drive} (P142; newest; CloudConvert).
%
Access to email and contacts were common concerns; for instance, P63 said, \enquote{As with any app that requires having access to send emails, I'm always worried about something unauthorized being sent} (oldest; Boomerang for Gmail). P61 (oldest; Quora) noted:
\begin{quote}
\small
I didn't know that they could see and download my contacts. That is a bit concerning because I don't know what they do with that data. 
\end{quote}
Participants were also concerned when they could not recall authorizing the access, e.g., \enquote{I don't remember authorizing this app to have access to my Google account} (P10, newest; Email - Edison Mail).
Additionally, when participants infrequently used an app but found out it still had access to their account, e.g., \enquote{I don't use it anymore and they still have access to my photos}  (P16; random; Chatbooks - Print Family Photos) and \enquote{I don't use the Google nest hub anymore, so it shouldn't have access to my full account} (P12; oldest; Google Nest Hub).
A number of participants complained that they had deleted the app from their device but it still had access to their accounts, as when P137 shared, \enquote{I have removed the app from my phone and I don't see why the app still has to have these permissions} (newest; Adidas Training), and P26 noted, \enquote{I was unaware these permissions were still on the app as I've deleted the app} (newest; Linkt).
Many participants stated that they were unconcerned by the app permissions (newest, $n = \num{62}$; oldest, $n = \num{44}$; random, $n = \num{41}$). 
The most common reasons for the lack of concern was the necessity of the permissions, and the limited access that the permissions provided.
For instance, P131 said \enquote{I don't think it has too many permissions and nothing seems unreasonable, so I'm okay with this} (random; Lumin PDF), and P158 stated \enquote{It is a game and both of the permissions are valid} (random; Starlost - Space Shooter).
Some were unconcerned because they trust the app, e.g., \enquote{I trust the service to keep my information secure} (P148; newest; Amazing Marvin).
For others, the trust originates with the company who makes the app, e.g., \enquote{I've never worried about Apple invading my privacy} (P189; random; macOS).
%

\subsection{Granting and Reviewing Account Access}
\figBeforeGrantingAccess
\paragraph{Considerations When Granting App/SSO Access.}
Google's documentation ~\cite{google-share-data-safely} advises that users consider the following five factors before granting a third-party access to their Google account: 
\begin{enumerate*}[label=(\roman*)]
\item security,
\item data use,
\item data deletion,
\item policy changes, and
\item data visibility.
\end{enumerate*} 

%
\changes{On a five-point agreement Likert-scale, we asked the participants with apps} if they considered these factors before granting a third-party app access to their Google account (\ref{appendix:main-survey:Q3}) and before using their Google account to sign into a service via SSO (\ref{appendix:main-survey:Q1}); those results are presented in \autoref{fig:beforeGrantingAccess-bar}. 
%

The most considered factor for third-party apps was the security of the app or website (security), in which \SI{83}{\percent} ($n = \num{118}$) \emph{Agree} or \emph{Strongly Agree}.
%
%
The next factor was how the app or website will use your data (data use) where \SI{78}{\percent} ($n = \num{111}$) \emph{Agree} or \emph{Strongly Agree}.
%
%
%
%
%
This was followed by who else can see your data on the app or website (data visibility), in which \SI{71}{\percent} ($n = \num{101}$) \emph{Agree} or \emph{Strongly Agree}.
%
%
Next was whether you can delete your data from the app or website (data deletion), where \SI{55}{\percent} ($n = \num{78}$) either \emph{Agree} or \emph{Strongly Agree}.
%
%
The least considered factor was whether the app or website will tell you if something changes (policy changes), where fewer than half ($n = \num{68}$; \SI{48}{\percent}) \emph{Agree} or \emph{Strongly Agree}.

%
\changes{Among participants with SSOs ($n=184$), the data shows they considered similar factors as those for third-party app account access.}
%
%
Again, the most considered factor was how secure is the app or website (security), in which \SI{83}{\percent} ($n = \num{153}$) \emph{Agree} or \emph{Strongly Agree}.
%
%
The next most considered factor was how the app or website will use your data (data use) where \SI{77}{\percent} ($n = \num{142}$) \emph{Agree} or \emph{Strongly Agree}.
%
%
This was followed by who else can see your data on the app or website (data visibility), in which \SI{72}{\percent} ($n = \num{133}$) \emph{Agree} or \emph{Strongly Agree}.
%
%
Next was whether you can delete your data from the app or website (data deletion), where \SI{57}{\percent} ($n = \num{105}$) either \emph{Agree} or \emph{Strongly Agree}.
%
%
The least considered factor was whether the app or website will tell you if something changes (policy changes), where fewer than half ($n = \num{88}$; \SI{48}{\percent}) \emph{Agree} or \emph{Strongly Agree}.
%

We used a Mann-Whitney U-test to compare each of the considerations, comparing SSO to third-party access (see top of \autoref{fig:beforeGrantingAccess-bar}). We did not find any significant differences, suggesting that participants view SSOs and third-party apps accessing their Google accounts in similar ways when determining if they should grant that access. More detail on SSO  considerations is in the next section.

%

In open-response questions, we asked participants to provide more details on  what they consider before granting a third-party app access to their Google account or use  SSOs~(\ref{appendix:pre-survey:S12},~\ref{appendix:pre-survey:S13}).
Participants often (Q12, $n = \num{42}$; Q13, $n = \num{43}$) responded that they consider what permissions the third-party would obtain, e.g., \enquote{What capabilities I was giving the third-party} (P362).
Security (Q12, $n = \num{33}$; Q13, $n = \num{23}$) and privacy (Q12, $n = \num{22}$; Q13, $n = \num{28}$) were common considerations. 
For instance P160 noted, \enquote{Whether it was secure and could I trust it,} and P283 added, \enquote{I'm always worried about my privacy anytime a app [sic] asks me for that information.}
%
Still, many participants (Q12, $n = \num{31}$; Q13, $n = \num{10}$) had no considerations. 
For instance, P24 shared, \enquote{It was a pop up so I didn't consider it much at all.}
One particularly interesting theme that emerged from open coding was the transfer of trust from the Google brand name to third-party apps and SSOs when participants considered granting access (Q12, $n = \num{7}$; Q13, $n = \num{4}$). That is, participants were more likely to trust a third-party service because it was using Google, and they trusted Google, irrespective of the nature of the third-party service.  For example,  P278 declared, \enquote{That it was okay since Google was allowing it,} and P297 added, \enquote{It must be okay since it partnered with Google.} 
%
P117 provided another example:
\begin{quote}
    \small
    I would consider nothing again, I probably put too much trust in Google and it's become a crutch at this point, I would easily allow it to be used in 3rd party situations.
\end{quote}
and P161:
\begin{quote}
    \small
    Nothing! Like before, I generally trust anything that leads to that Google SSO page.
\end{quote}
%

%
%

When considering granting SSO access~(\ref{appendix:pre-survey:S8}),
many participants ($n = \num{55}$) considered ease of use and convenience before signing in, and a common theme was the ease of reusing their Google account login credentials versus creating a new account on a third-party app or website, e.g., \enquote{It is easier and more convenient than making a brand new account to some third party website} (P21). 
There were also many ($n = \num{42}$) who were unconcerned and had few considerations, such as P99 who said,\enquote{I didn't consider much, I use Google sign in pretty regularly,} and P117 who stated, \enquote{I didn't consider much, my Google account has always been good to me and offered ease of use with many things.}
Participants ($n = \num{27}$) also shared concerns such as what information would be accessible to the third-party app or service; for instance, P90 responded, \enquote{If that service has access to information associated with my Google Account,} and P91 replied, \enquote{I considered what that app would have access to if I signed in through there.}
Security ($n = \num{29}$) and privacy ($n = \num{13}$) were common considerations. 
For example P80 shared, \enquote{I was worried about security of my Google Account,} and P271 added, \enquote{The private data that this application was going to read, in other words, I worry about my privacy.}
Some participants ($n = \num{12}$) described a trade-off between information sharing and convenience. For instance, P16 shared, \enquote{The effort of making a new account versus sharing my google info.}
Trust of the website or service being signed into is also a common theme ($n = \num{26}$), e.g., \enquote{I consider if the website is trustworthy and if I can trust signing in using my Gmail account in their website} (P68).
%

\paragraph{Reasons For Authorizing Account Access.}
%
We asked participants what was the purpose of allowing third-party  access~(\ref{appendix:pre-survey:S11}). 
The purpose for many ($n = \num{45}$) participants was the utility that the app provided, such as email and contact management, file transfer, and synchronizing data between devices. 
For instance, P37 explained, \enquote{I gave a ringtone app access to my contacts and messages so that it can change the notification/ringtone sounds} and P210 shared that the apps purpose was \enquote{Allowing me access to multiple e-mails from the same app on my Windows computer.}
Another popular ($n = \num{38}$) purpose was calendar management, e.g., \enquote{Zoom, to allow it to add meetings to my calendar} P25.
Gaming was a common ($n = \num{38}$) purpose. For example, P355 responded:
\begin{quote}
    \small
    I allowed access to my Google account for a lot of games, because I can get achievements to be displayed on my account with Google, and also it is usually an easy way to recover or save data between devices.
\end{quote}
Some participants ($n = \num{15}$) just could not recall the purpose, like P404 who shared, \enquote{I don't really remember but I do remember encountering this type of screen in the past.}
%

\figHowOftenReviewAppSSO
\paragraph{User Review of Apps With Account Access.}
We find that participants rarely or never review the services they can sign into using their Google account or the apps that have access to their Google account. 
%
%
We asked participants how often they review the services they can sign into using their Google account (\ref{appendix:main-survey:Q2}). A majority ($n = \num{106}$; \SI{58}{\percent}) \emph{Rarely} review their services, \SI{20}{\percent} ($n = \num{37}$) \emph{Never} review, and \SI{16}{\percent} ($n = \num{29}$) review them \emph{Monthly} or \emph{Yearly}.
%
%
\changes{Among participants with apps, we also asked} how often they review the services that have access to their Google account (\ref{appendix:main-survey:Q4}). Again, a majority ($n = \num{80}$; \SI{56}{\percent}) \emph{Rarely} review their services, \SI{24}{\percent} ($n = \num{35}$) \emph{Never} review, and \SI{18}{\percent} ($n = \num{26}$) review them \emph{Monthly} or \emph{Yearly}.
\autoref{fig:howOftenReviewAppSSO-bar} shows the full results of these questions.

%
%
%

For each of their newest, oldest and random app, we also asked participants if they would like to change which parts of their Google account that the third-party app can access (\ref{appendix:main-survey:Q10}).
\SI{52}{\percent} ($n = 74$) of participants \emph{Agree} or \emph{Strongly agree} they want to change which parts of their Google account are accessible for at least one of their apps.
The most agreement for changing access was for the random app,  in which 30\% indicated they \emph{Agree} (15\%; $n = \num{13}$) or even \emph{Strongly agree} (15\%; $n = \num{13}$). 
This was followed by the newest app, where 21\% ($n = 25$) \emph{Agree} and 9\% ($n = 11$) \emph{Strongly agree}. 
The oldest app had the least agreement for a change in access with only 15\% ($n = 22$) \emph{Agree} and 7\% ($n = 10$) \emph{Strongly agree}.
%
%

%
%

\paragraph{Keeping or Removing Apps.} 
For each of the specific apps shown---newest, oldest, and randomly chosen---we asked participant if they would like to keep or remove the app or are unsure about what to do (\ref{appendix:main-survey:Q5}).
\SI{43}{\percent} ($n = 62$) of participants with third-party apps wanted to remove at least one of those apps.
Due to private data aggregation during the survey, we only linked the keep/remove preferences for the newest, oldest, and random app that were specifically reviewed by the participants. 
For their newest app, \SI{56}{\percent} ($n = \num{65}$) of the participants said they want to keep it, \SI{30}{\percent} ($n = \num{35}$) chose to remove, and \SI{15}{\percent} ($n = \num{17}$) answered unsure.
Many more participants (\SI{76}{\percent}; $n = \num{108}$) responded to keep their oldest app. While \SI{20}{\percent} ($n = \num{28}$) wanted to remove, and \SI{5}{\percent} ($n = \num{7}$) were unsure.
%
%
\SI{60}{\percent} ($n = \num{51}$) of participants wanted to keep their randomly selected app, 27\% ($n = \num{23}$) answered to remove, and 13\% (n = $\num{11}$) were unsure. However, a Kruskal-Wallace test found no significant differences between the three apps shown ($H=0.49,p=0.77$).
Full results in  \autoref{fig:awareRecallKeep-bar}.

%
%
%
%

%

\begin{table}[tbp]
\centering
\caption[Remove app regression analysis]{\label{tab:remove-app-regression}
Binomial logistic model to describe which factors influenced the preference to remove an app (\emph{Remove} responses to question \ref{appendix:main-survey:Q5}).
The  Aldrich-Nelson pseudo  $R^2 = \num{0.52}$.
}
{
\small
\begin{tabular*}{\columnwidth}{
@{}
l 
@{\extracolsep{\fill}}
S[table-format = -1.2]
S[table-format = 1.2]
@{\extracolsep{4pt}}
S[table-format = 1.2, parse-numbers = false, parse-units = false]
@{\extracolsep{4pt}}H@{\extracolsep{\fill}}
S[table-format = <1.3]
@{\extracolsep{1pt}}
l
}
    \toprule
    {\textbf{Factor}} & {\textbf{Est.}} & {\textbf{OR}} & {\textbf{Error}} & {\textbf{z value}} & {\textbf{Pr(\textgreater\textbar z\textbar)}} & \\ 
    \midrule
(Intercept) & -1.02 & 0.36 & 0.68 & -1.50 & 0.134 &  \\ 
$\text{Participant's newest app}$ & 0.22 & 1.24 & 0.53 & 0.40 & 0.687 &  \\
$\text{Participant's oldest app}$ & -0.07 & 0.93 & 0.56 & -0.12 & 0.904 &  \\ 
$\text{Recall} = Yes$ & 0.92 & 2.50 & 0.68 & 1.34 & 0.179 &  \\ 
$\text{Aware app permissions} = Yes$ & -1.15 & 0.32 & 0.61 & -1.90 & 0.058 & . \\ 
$\text{Last use} \in \{\textit{day, week, month}\}$ & -1.76 & 0.17 & 0.49 & -3.57 & <0.001 & *** \\ 
$\text{Access benef.} \in \{\textit{Agr., Str. Ag.}\}$ & -1.80 & 0.17 & 0.55 & -3.29 & 0.001 & ** \\ 
$\text{Access conc.} \in \{\textit{Agr., Str. Ag.}\}$ & 0.43 & 1.53 & 0.56 & 0.77 & 0.443 &  \\ 
$\text{Access change} \in \{\textit{Agr., Str. Ag.}\}$ & 1.75 & 5.76 & 0.54 & 3.25 & 0.001 & ** \\ 
$\text{\# of permissions} > \text{median}$ & 0.05 & 1.05 & 0.44 & 0.11 & 0.909 &  \\ 
$\text{Time since install} < \text{3 months}$ & 0.84 & 2.32 & 0.54 & 1.54 & 0.123 &  \\
$\text{Time since install} > \text{2 years}$ & 0.26 & 1.30 & 0.66 & 0.40 & 0.689 &  \\
  \bottomrule
\end{tabular*}
}
\footnotesize
\textbf{Signif. codes:} $\text{***}~\widehat{=} < 0.001$; $\text{**}~\widehat{=} <0.01$; $\text{*}~\widehat{=} <0.05$; $\cdot~\widehat{=} <0.1$
\end{table}

{}

%
We performed logistic regression to determine factors that would lead a participant to remove a third-party app access from their Google account (see \autoref{tab:remove-app-regression}). We controlled for repeated measures by adding random intercepts to the model, as each participant provided up to three apps they wished to keep or remove.  
We found a significant correlation with participants who want to change which parts of their Google account the app can access ($\beta=1.75, OR=5.76, p=0.001$), where those participants were $5.8\times$ more likely to want to remove the app access.  
We found a significant correlation with participants who have used the app within the last month ($\beta=-1.76, OR=0.17, p=<0.001$), and those participants are $5.9\times$ more likely to want to keep the app access.
We also found significant correlation with participants who found the app beneficial ($\beta=-1.80, OR=0.17, p=0.001$), where they too are $5.9\times$ more likely to want to keep the app access. 
These findings suggest the importance of app usage frequency in account access authorization.

\subsection{Reflection and Features}




\paragraph{Understanding Account Linking and Access.}
\changes{We directed all participants in the second survey ($n=214$) to explore their \enquote{Apps with access to your account} page. We} then asked whether the page helps them to better understand which third-party apps and websites are linked to their Google account~(\ref{appendix:main-survey:Q15}); nearly all participants ($n = \num{204}$; \SI{95}{\percent}) agreed that it help.
%
And when asked if the page helps to better understand which parts of their Google account third-party apps can access~(\ref{appendix:main-survey:Q16}), again nearly all participants ($n = \num{198}$; \SI{93}{\percent}) agreed.
This suggests that the management page has the potential for being an important tool.

%

\figChangeSettingsReviewApps


\paragraph{Change Settings and Review Apps.}
We asked participants to indicate whether they intend to change any settings after seeing their \enquote{Apps with access to your account} page~(\ref{appendix:main-survey:Q17}), and roughly half ($n = \num{105}$) affirmed that they would change settings.
%
%
Over \SI{70}{\percent} ($n = \num{152}$) indicated they plan to review third-party apps in six months~(\ref{appendix:main-survey:Q18}).
%
Refer to \autoref{fig:changeSettingsReviewApps-bar} to see the full results of these questions. 
%

If a participant affirmed they would change a setting, we asked them to tell us which settings they would change (\ref{appendix:main-survey:Q19}).
Many participants ($n = \num{76}$) wanted to remove access from one or more apps.
They often ($n = \num{41}$) mentioned unused apps; for instance, P23 said:
\begin{quote}
    \small
    I would definitely remove many of the apps that I do not use anymore. They absolutely do not need to be linked to my Google account anymore.
\end{quote}
Participants also wished to remove apps they no longer recalled authorizing, e.g.,
\enquote{There are apps I do not use and do not recall allowing access to my Google account} (P24).
Some participants ($n = \num{27}$) wanted to change specific permissions access, something not allowable with the current interface. The most common permissions mentioned included contacts, account info, delete or modify files, and \enquote{unnecessary permissions.} For example, P189 said, \enquote{I would remove Dropbox's access to my contacts,} and P192 noted, \enquote{\dots and maybe restrict Streak's access to my Google Drive.}
Participants ($n = \num{8}$) mentioned protecting their personal information as a reason for the settings change, e.g., \enquote{I would limit the amount of information different apps can view - especially with my personal information} (P145).
%
 
%
If a participant affirmed they planned to review third-party apps in six months, we asked them to tell us what they would look for~(\ref{appendix:main-survey:Q20}).
Many participants ($n = \num{43}$) would look for apps that they no longer use, such as P137, who replied \enquote{I would look for apps that I no longer use and if they still have access to my account and to what parts of my account.}
Some participants ($n = \num{22}$) said they would look for new apps that have access to their account, e.g., \enquote{If any new apps are listed that I don't remember approving} (P169). 
Others ($n = \num{22}$) wanted to review changes to the list of apps with access to their Google account. For instance, P138 explained:
\begin{quote}
    \small
    I would look for any unexpected or new third parties and just make sure that all of them have very limited access to my personal information. I'd maybe continue to review this page over time to make sure nothing has changed.
\end{quote}
Some wanted to review changes to the existing apps, like P180 who shared, \enquote{Check to see if anything has changed like they got more access without saying anything.}
Participants ($n = \num{8}$) also wanted to review how much access was allowed to third-party apps. For example, P210  stated:
\begin{quote}
    \small
    I would want to know exactly what those apps have access to. Actually wish there was more information there. Seems like it is a little generic.
\end{quote}

\paragraph{New Features and Design Changes.}
In an open-ended response question, we asked what new features participants would like to add to the ``Apps with access to your account'' page~(\ref{appendix:main-survey:Q21}).
Many participants ($n = \num{61}$) wanted the page to display more detailed information and for the account access to be more transparent, e.g, \enquote{An option to have even more detail of what was exactly being accessed} (P68). 
A common ($n = \num{19}$) request was an app usage or data access log, like when P17 replied: 
\begin{quote}
    \small
    I think it would be useful for them to show when I last used an app and when the app last used my data, to see if the app is using my data even when I have not used the app in a while.
\end{quote}
Another request ($n = \num{12}$) was detailed permission explanations, for instance when P174 demanded,
\begin{quote}
    \small
    DETAILED information about what is available to third party apps. Not just \enquote{access to your personal information,} but \enquote{access to: your full name, age, date of birth, street address.}
\end{quote}
Permissions level control is also desired, e.g., \enquote{The ability to remove certain accesses that I don't think the app needs} (P143), and some requested access timeouts, e.g., \enquote{Auto removal of apps after 3 months of no use} (P156).

\figRemindAndReapprove
We asked participants how often they would like to be reminded if Google were to provide an email reminder to review their \enquote{Apps with access to your account} page~(\ref{appendix:main-survey:Q22}). 
Most participants ($n = \num{117}$; \SI{55}{\percent}) responded \emph{Once a month}, another \SI{37}{\percent} ($n = \num{79}$) want a reminder \emph{Once a year}.

%

 \paragraph{Reapproving Apps.}
Additionally, we asked participants if Google required them to reapprove the third-party apps with access to their account,  how often would they want to provide reapproval~(\ref{appendix:main-survey:Q22}). 
A majority of participants ($n = \num{127}$; \SI{59}{\percent}) want to reapprove apps \emph{Once a year}, while \SI{29}{\percent} ($n = \num{63}$) want a reminder \emph{Once a month}.
Please see \autoref{fig:remindAndReapprove-bar} for the full results of these questions. 

\figApproveEveryBlockData
When we asked participants if they would want third-party apps to seek approval each time they access their Google account~(\ref{appendix:main-survey:Q25}) over half ($n = \num{113}$; \SI{53}{\percent}) agreed they want third-party apps to seek approval each time they access their Google account, while \SI{27}{\percent} ($n = \num{58}$) disagreed.
We also asked participants if they want to designate specific data (e.g. certain emails, individual contacts, particular calendar events) as private and inaccessible to third-party apps~(\ref{appendix:main-survey:Q26}), and most participants ($n = \num{189}$; \SI{89}{\percent}) agreed, few ($n = \num{3}$; \SI{1}{\percent}) disagreed.
Please refer to \autoref{fig:approveEveryBlockData-bar} for the full results.
%

\section{Discussion and Conclusion}\label{sec:discussion}

In this section we offer conclusions and recommendations for managing third-party apps and SSOs that access users' Google accounts. We first discuss issues with handling disused third-party access and the potential to have cascading removal. Next, we focus on issues with transference of trust from Google to third-party apps, and finally, we offer some design improvement for \enquote{Apps with access to your account} based on our observations and qualitative feedback. 

\paragraph{Handling Stale Account Access.}
This research identifies the need for limiting account access for unused, forgotten, or removed third-party apps. Participants either had not recently used or were unsure when they had last used one in five third-party apps in our study. Additionally, most participants (~80\%) rarely or never reviewed the third-party account access. At the same time, nearly half of participants wanted to modify access after the study and over 70\%   wanted to do so within six-months.  Most participants indicated they would remove unused apps or apps they no longer recall authorizing, 
but the current interface for reviewing third-party apps does not provide a record of last use.

Participants also expressed strong support (95\%) for email reminders to review their third-party apps at least once a year. Moreover, participants highly favored  (91\%) a requirement to reapprove account access at least once a year. Google could send email reminders to all users to review \enquote{Apps with access to your account} with additional details provided regarding unused third-party apps. Review request could also be more directed, perhaps asking users to review a single app at a time so as not to overburden. 
%
%
Another possibility is to auto-expire apps that have not been recently used or did not recently make an API call. Facebook implemented a similar policy following the Cambridge Analytica scandal~\cite{fb-user-data-privacy}, to expire app access periodically and then require users to reauthorize.

\paragraph{Cascading Removal.}
Participants were surprised to find apps in their \enquote{Apps with access to your account} page that they believed were removed in other places. For example, many smartphone applications leverage a Google Account for functionality and authentication, and in these case, participants who removed that smartphone app believed they also remove the third-party app access.

While expiring third-party apps, \changes{see above}, would help address this issue, cascading removal could provide additional support for users who attempt to manage access to their Google account. A third-party app, or the user's account, can send a notification when a secondary application that leverages the third-party app is removed from a user device, which in turn allows the user to 
deactivate the third-party app access.

\paragraph{All Or Nothing Permissions.}
Users must approve {\em all} permission requests for third-party apps or none, and many of permission requests may not be fully understandable, contextualized, or presented at the time of the access request. After granting access, the \enquote{Apps with access to your account} offers no permission-level control to limit account access; the only option is to completely remove access.

Even without fine-grain controls, most participants in our study find most permissions to be necessary for the functionality of their apps; however, they consider some permissions unnecessary and concerning, especially those that allow access to view personal information that may include data beyond just name and email address. Our qualitative data shows that this practice is forcing users to consider a tradeoff between their personal privacy and the benefits of third-party apps or the convenience of SSOs.

Furthermore, when prompted to describe new features, participants stated a desire for permission level controls, and we recognize such a feature could have negative impacts on the functionality of third-party apps. However, we recommend that Google provide permissions level control for users' personal data, for instance: \enquote{See your gender,} \enquote{See your age group,} \enquote{View your street addresses,} \enquote{See and download your personal phone numbers,} \enquote{See your personal info,} and \enquote{Associate you with your personal info on Google,} which likely are not used for app functionality. 

\paragraph{Trust Transference.}
Google is a highly ranked and trusted brand~\cite{google-trusted-company}, \changes{so trust} in Google transfers to Google products and services, including third parties' apps that access Google accounts and services. We observed this in many open-ended responses from participants that assumed (incorrectly) an implied partnership between Google and third-party apps. 

To Google's credit, it attempts to  mitigate this effect by noting that users should \enquote{Make sure you trust [\emph{third-party app name}],} on the access authorization prompt. However, this does not appear to be a sufficient intervention, and more so, it places a potential undue burden on users to be able to properly differentiate. This further suggests that better messaging, in the form of nudges or reminders to review third-party apps, or other automated mechanisms should be put in place to assist users to better manage apps and services that access their Google account. 

\paragraph{Improving App Transparency Tool.}
We also note that there are number of possible design improvements to the existing \enquote{Apps with access to your account} page based on participant recommendations offered during the study and our findings. Nearly 90\% of participants suggested that Google should provide users with the ability to designate specific account data as private and inaccessible to third-party apps.
For instance: certain emails, individual contacts, and particular calendar events. Moreover, adding a recent activity log that includes data access details would allow users to determine the frequency with which they use a third-party app and to determine if it is accessing data while appearing inactive. 

Improvements to the permission descriptive text could also include specific details about which parts of the users' Google account data are accessible.  For example, instead of \enquote{See your personal info,} the personal info available to see should be enumerated, e.g., full name, email address, age, street address, and profile picture. Contextual information could also be included, such as identifying the reasons a third-party app is requesting a certain permission. For instance, \enquote{Zoom has access to your Google calendar to schedule and modify Zoom meeting times,} instead of simply noting it has access to the calendar.


\clearpage
\subsection*{Acknowledgements}
\noindent
We thank Florian Farke for providing R script examples of high quality figures and tables. We thank the research team members of The George Washington University Usable Security and Privacy Lab for their valuable feedback on the survey questionnaires. We also thank Michael Lack and Dan Hallenbeck of Two Six Technologies for their support and feedback on the initial investigation into G Suite apps that prompted this research.

\appendix
\section{Appendix}\label{sec:appendix}

\setlength{\columnsep}{0pt}
\setlength{\multicolsep}{0pt}

\subsection{First Survey}\label{appendix:pre-survey}
\begin{footnotesize}

\textbf{Please read the following instructions carefully:}
\label{appendix:pre-survey:survey-instructions}
\begin{itemize}[nosep]
    \item Take your time in reading and answering the questions.
    \item Answer the questions as accurately as possible.
\end{itemize}

\begin{questions}[label=\textbf{Q$\mathbf{{}_1}$\arabic*}]

    \item Do you have a Gmail (Google) account?
    \label{appendix:pre-survey:S1}
    \begin{multicols}{2}
        \begin{answers}
            \item Yes 
            \item No
        \end{answers}
    \end{multicols}





\noindent\textbf{Privacy Notice:}
We do not transmit your email address to our server as part of this study, and we will not be able to tie your email address to any results or analysis. All uses of your email address are local to your browser. The researchers will never see your email address. At no time do the researchers have access to your Google account.


\emph{[\ref{appendix:pre-survey:S2} through \ref{appendix:pre-survey:S13} are shown if \ref{appendix:pre-survey:S1} is \enquote{Yes}]}

    \item Do you use \{\emph{participant email address}\} as your primary Gmail (Google) account?
    \label{appendix:pre-survey:S2}
    \begin{answers}
        \item Yes 
        \item No, I use a different Gmail (Google) as my primary account
    \end{answers}

    \item Whose Gmail address is it?
    \label{appendix:pre-survey:S3}
    \begin{answers}
        \item It is my own account. I have sole ownership of this account. 
        \item It is an institutional account. A business, school, or organization gave it to me.
        \item It is my shared account. I share the account with someone else (e.g., a partner or family member).
        \item It is someone else’s account. Someone else has sole ownership of this account.
    \end{answers}

    \item How long have you had this Gmail address as your primary Gmail (Google) account?
    \label{appendix:pre-survey:S4}
    \begin{multicols}{2}
        \begin{answers}
            \item Less than 1 month 
            \item Less than or about 1 year
            \item Less than or about 5 years
            \item Less than or about 10 years
            \item More than 10 years
            \item Unsure
        \end{answers}
    \end{multicols}

    \item What is the color of a red ball?
    \label{appendix:pre-survey:S5}
    \begin{multicols}{4}
        \begin{answers}
            \item Red
            \item Round
            \item Blue
            \item Square
        \end{answers}
    \end{multicols}
%
\textbf{Use your Google Account to sign in to other apps or services}
\begin{itemize}[nosep]
    \item You can use your Google Account to sign in to third-party apps and services. 
    \item You won't have to remember individual usernames and passwords for each account.
    \item Third-party apps and services are created by companies or developers that aren't Google.
\end{itemize}

    \item Do you recall ever using your Google Account to sign in to third-party apps or services as described above?
    \label{appendix:pre-survey:S6}
    \begin{multicols}{3}
        \begin{answers}
            \item Yes
            \item No
            \item Unsure
        \end{answers}
    \end{multicols}
    
    \item Thinking about the last time you used your Google Account to sign into a third-party app or service, what app or service did you use your Google Account to sign into?
    \emph{[Shown if \ref{appendix:pre-survey:S6} is \enquote{Yes}]}
    \label{appendix:pre-survey:S7} (short answer)

    \item Thinking about the last time you used your Google Account to sign into a third-party app or service, what did you consider before signing in using your Google Account?
    \emph{[Shown if \ref{appendix:pre-survey:S6} is \enquote{Yes}]}
    \label{appendix:pre-survey:S8} (short answer)
    
    \item If you were given the option to use your Google account to sign into a third-party app or service, what would you consider before using this feature?
    \emph{[Shown if \ref{appendix:pre-survey:S6} is \enquote{No} or \enquote{Unsure}]}
    \label{appendix:pre-survey:S9} (short answer)

\textbf{Manage third-party apps \& services with access to your Google Account}
\begin{itemize}[nosep]
    \item Google lets you give third-party apps and services access to different parts of your Google Account.  
    \item Third-party apps and services are created by companies or developers that aren’t Google.
    \item For example, you may download an app that helps you schedule workouts with friends. This app may request access to your Google Calendar and Contacts to suggest times and friends for you to meet up with.  
\end{itemize}
    
    \item Do you recall ever granting a third-party app access to your Google Account as described above?
    \label{appendix:pre-survey:S10}
    \begin{multicols}{3}
        \begin{answers}
            \item Yes
            \item No
            \item Unsure
        \end{answers}
    \end{multicols}

    \item Thinking about the last time you granted a third-party app access to your Google Account, what was the purpose of allowing that access?
    \emph{[Shown if \ref{appendix:pre-survey:S10} is \enquote{Yes}]}
    \label{appendix:pre-survey:S11} (short answer)
    
    \item Thinking about the last time you granted a third-party app access to your Google Account, what did you consider before granting a third-party app access to your Google account?
    \emph{[Shown if \ref{appendix:pre-survey:S10} is \enquote{Yes}]}
    \label{appendix:pre-survey:S12} (short answer)
    
    \item If you were given the option to grant a third-party app access to your Google Account, what would you consider before granting access?
    \emph{[Shown if \ref{appendix:pre-survey:S10} is \enquote{No} or \enquote{Unsure}]}
    \label{appendix:pre-survey:S13} (short answer)

\emph{[\ref{appendix:pre-survey:S14} through \ref{appendix:pre-survey:S23} are shown if \ref{appendix:pre-survey:S1} is \enquote{No}]}


\textbf{Use an existing online account to sign in to other apps or services}
\begin{itemize}[nosep]
    \item You can use your existing account on various online platforms (e.g., Facebook, Google, Apple, etc.) to sign in to third-party apps and services.  
    \item You won't have to remember individual usernames and passwords for each account.
    \item Third-party apps and services are created by companies or developers that aren’t part of those platforms.
\end{itemize}

    \item Do you recall ever using an existing online account to sign in to a third-party app or service as described above?
    \label{appendix:pre-survey:S14}
    \begin{multicols}{3}
        \begin{answers}
            \item Yes
            \item No
            \item Unsure
        \end{answers}
    \end{multicols}

    \item Thinking about the last time you used an existing online account to sign into a third-party app or service, what app or service did you sign into?
    \emph{[Shown if \ref{appendix:pre-survey:S14} is \enquote{Yes}]}
    \label{appendix:pre-survey:S15} (short answer)
    
    \item Thinking about the last time you used an existing online account to sign into a third-party app or service, what online account did you use?
    \emph{[Shown if \ref{appendix:pre-survey:S14} is \enquote{Yes}]}
    \label{appendix:pre-survey:S16} (short answer)
    
    \item Thinking about the last time you used an existing online account to sign into a third-party app or service, what did you consider before signing in using that existing online account?
    \emph{[Shown if \ref{appendix:pre-survey:S14} is \enquote{Yes}]}
    \label{appendix:pre-survey:S17} (short answer)
    
    \item If you were given the option to use an existing online account to sign into a third-party app or service, what online account would you use?
    \emph{[Shown if \ref{appendix:pre-survey:S14} is \enquote{No} or \enquote{Unsure}]}
    \label{appendix:pre-survey:S18} (short answer)
    
    \item If you were given the option to use an existing online account to sign into a third-party app or service, what would you consider before using this feature?
    \emph{[Shown if \ref{appendix:pre-survey:S14} is \enquote{No} or \enquote{Unsure}]}
    \label{appendix:pre-survey:S19} (short answer)


\textbf{Manage third-party apps \& services with access to your online Account}
\begin{itemize}[nosep]
    \item Online platforms let you give third-party apps and services access to different parts of your accounts on those platforms.  
    \item Third-party apps and services are created by companies or developers that aren't part of those platforms.
    \item For example, you may download an app that helps you schedule workouts with friends. This app may request access to your online calendar (e.g., Google Calendar, Apple iCloud Calendar, Microsoft Outlook, etc.) and contact list to suggest times and friends for you to meet up with.  
\end{itemize}

\clearpage

    \item Do you recall ever granting a third-party app access to one of your online accounts as described above?
    \label{appendix:pre-survey:S20}
    \begin{multicols}{3}
        \begin{answers}
            \item Yes
            \item No
            \item Unsure
        \end{answers}
    \end{multicols}

    \item Thinking about the last time you granted a third-party app access to one of your online accounts, what was the purpose of allowing that access?
    \emph{[Shown if \ref{appendix:pre-survey:S20} is \enquote{Yes}]}
    \label{appendix:pre-survey:S21} (short answer)

    \item Thinking about the last time you granted a third-party app access to one of your online accounts, what did you consider before granting a third-party app access to one of your online accounts?
    \emph{[Shown if \ref{appendix:pre-survey:S20} is \enquote{Yes}]}
    \label{appendix:pre-survey:S22} (short answer)

    \item If you were given the option to grant a third-party app access to one of your online accounts, what would you consider before granting access?
    \emph{[Shown if \ref{appendix:pre-survey:S20} is \enquote{No} or \enquote{Unsure}]}
    \label{appendix:pre-survey:S23} (short answer)

    \item[] \emph{These questions were followed by the 8 IUIPC items as described by Malhotra~\etal~\cite{malhotra-2004-iuipc} and  Gro{\ss}~\cite{gross2021validity}.}
    
\end{questions}

\begin{questions}[label=\textbf{D\arabic*}]

    \item What is your gender?
    \label{appendix:pre-survey:D1}
    \begin{multicols}{2}
        \begin{answers}
            \item Woman
            \item Man
            \item Non-binary
            \item Prefer not to disclose
            \item Prefer to self-describe
        \end{answers}
    \end{multicols}
    
    \item What is your age?
    \label{appendix:pre-survey:D2}
    \begin{multicols}{4}
        \begin{answers}
            \item 18 -- 24
            \item 25 -- 34
            \item 35 -- 44
            \item 45 -- 54
            \item 55 -- 64
            \item 65 or older
            \item Prefer not to disclose
        \end{answers}
    \end{multicols}
    
    \item What is the highest degree or level of school you have completed?
    \label{appendix:pre-survey:D3}
    \begin{answers}
        \item No schooling completed
        \item Some high school
        \item High school
        \item Some college credit, no degree
        \item Trade, technical, or vocational training
        \item Associate degree
        \item Bachelor's degree
        \item Master's degree
        \item Professional degree
        \item Doctorate degree
        \item Prefer not to disclose
    \end{answers}

    \item Which of the following best describes your educational background or job field?
    \label{appendix:pre-survey:D4}
    \begin{answers}
        \item I have an education in, or work in, the field of computer science, computer engineering or IT.
        \item I do not have an education in, nor do I work in, the field of computer science, computer engineering or IT.
        \item Prefer not to disclose
    \end{answers}
    
\end{questions}

\end{footnotesize}


\subsection{Second Survey}\label{appendix:main-survey}

\begin{footnotesize}






%
%

\textbf{About Account Linking}
\begin{itemize}[nosep]
    \item Users of online services can often link their accounts to products made by other companies, which can operate on behalf of the user's account. For example, a user can link their Spotify account (i.e., a music streaming service) to their Amazon Alexa (i.e., a voice-controlled smart speaker) in order to play their favorite music.
    \item Your Gmail account, as a Google account, can also be linked in this way. In the following questions, we ask about how your Google account can similarly be linked to other (non-Google) services. Later we will use a web browser extension to do further analysis of how you may (or may not) link your Google account.
\end{itemize}





%

\textbf{Explore Apps With Access To Your Account}
\begin{itemize}[nosep]
    \item In the next part of the study, we will ask you to explore Google's ``Apps with access to your account'' page for your Google account. 
    \item You will have an opportunity to interact with your Google's ``Apps with access to your account'' page for \textbf{one minute} and will then be returned to the survey.
    \item We have \textbf{disabled clicking} for various links and buttons on the Google's ``Apps with access to your account''page to help you focus better on the access that third party applications have to your Google account.
\end{itemize}

\begin{center}
\emph{Participants explored their Apps with access to your account page.}
\end{center}



\textbf{Use your Google Account to sign in to other apps or services}
\begin{itemize}[nosep]
    \item You can use your Google Account to sign in to third-party apps and services. You won't have to remember individual usernames and passwords for each account.
    \item Third-party apps and services are created by companies or developers that aren't Google.
\end{itemize}

\begin{center}
\emph{Participants view a list of SSOs authorized to their Google account.}
\end{center}

\begin{questions}[label=\textbf{Q$\mathbf{{}_2}$\arabic*}]

    \item You have [\emph{n}] third-party apps and services that you can \textbf{sign into using your Google account}. Before using my Google account to sign into a website or third-party app, I consider:
    \label{appendix:main-survey:Q1}
    
    \begin{tiny}
    \hspace*{\fill}Strongly disagree Disagree Neither agree nor disagree Agree Strongly agree \\
    \end{tiny}
    \begin{scriptsize}
    \begin{tabularx}{\linewidth}{@{}l@{}C{1}@{}C{1}@{}C{1}@{}C{1}@{}C{1}@{}}  
        \toprule
        ...how secure the app or website is.& $\bigcirc$ & $\bigcirc$ & $\bigcirc$ & $\bigcirc$ & $\bigcirc$ \\
        ...how the app or website will use your data.& $\bigcirc$ & $\bigcirc$ & $\bigcirc$ & $\bigcirc$ & $\bigcirc$ \\
        ...whether you can delete your data from the app or website.& $\bigcirc$ & $\bigcirc$ & $\bigcirc$ & $\bigcirc$ & $\bigcirc$ \\
        ...whether the app or website will tell you if something changes.& $\bigcirc$ & $\bigcirc$ & $\bigcirc$ & $\bigcirc$ & $\bigcirc$ \\
        ...who else can see your data on the app or website.& $\bigcirc$ & $\bigcirc$ & $\bigcirc$ & $\bigcirc$ & $\bigcirc$ \\
        \bottomrule
    \end{tabularx}
    \end{scriptsize}

    \item How often do you review what services you can sign into using your Google account?
    \label{appendix:main-survey:Q2}
    \begin{multicols}{2}
        \begin{answers}
            \item Never
            \item Rarely
            \item Daily
            \item Weekly
            \item Monthly
            \item Yearly
        \end{answers}
    \end{multicols}
    
\textbf{Manage third-party apps \& services with access to your Google Account}
\begin{itemize}[nosep]
    \item Google lets you give third-party apps and services access to different parts of your Google Account. 
    \item Third-party apps and services are created by companies or developers that aren't Google.
    \item For example, you may download an app that helps you schedule workouts with friends. This app may request access to your Google Calendar and Contacts to suggest times and friends for you to meet up with.
\end{itemize}

\begin{center}
\emph{Participants view third-party apps with access to their Google account.}
\end{center}

    \item You have [\emph{n}] third-party apps and services that \textbf{have access to your Google account data}. Before granting a website or third-party app access to my Google account, I consider:
    \label{appendix:main-survey:Q3}
    
    \begin{tiny}
    \hspace*{\fill}Strongly disagree | Disagree | Neither agree nor disagree | Agree | Strongly agree \\
    \end{tiny}
    \begin{scriptsize}
    \begin{tabularx}{\linewidth}{@{}l@{}C{1}@{}C{1}@{}C{1}@{}C{1}@{}C{1}@{}} 
        \toprule
        ...how secure the app or website is.& $\bigcirc$ & $\bigcirc$ & $\bigcirc$ & $\bigcirc$ & $\bigcirc$ \\
        ...how the app or website will use your data.& $\bigcirc$ & $\bigcirc$ & $\bigcirc$ & $\bigcirc$ & $\bigcirc$ \\
        ...whether you can delete your data from the app or website.& $\bigcirc$ & $\bigcirc$ & $\bigcirc$ & $\bigcirc$ & $\bigcirc$ \\
        ...whether the app or website will tell you if something changes.& $\bigcirc$ & $\bigcirc$ & $\bigcirc$ & $\bigcirc$ & $\bigcirc$ \\
        ...who else can see your data on the app or website.& $\bigcirc$ & $\bigcirc$ & $\bigcirc$ & $\bigcirc$ & $\bigcirc$ \\
        ...what parts of your account the app or website can access.& $\bigcirc$ & $\bigcirc$ & $\bigcirc$ & $\bigcirc$ & $\bigcirc$ \\
        \bottomrule
    \end{tabularx}
    \end{scriptsize}

    \item How often do you review what services have access to your Google account?
    \label{appendix:main-survey:Q4}
    \begin{multicols}{5}
        \begin{answers}
            \item Never
            \item Rarely
            \item Daily
            \item Weekly
            \item Monthly
            \item Yearly
        \end{answers}
    \end{multicols}

    \item The following apps are authorized to access various parts of your Google account. Which of these apps would you prefer to keep on your account? Which would you prefer to remove?
    \label{appendix:main-survey:Q5}
    \begin{tabular*}{\linewidth}{@{}l@{\extracolsep{\fill}}*{6}{c@{}}}
        \toprule
                       & Keep  & Remove  & Unsure     \\
        \midrule
        \emph{[App 1]} & $\bigcirc$ & $\bigcirc$ & $\bigcirc$ \\
        \emph{[App 2]} & $\bigcirc$ & $\bigcirc$ & $\bigcirc$ \\
        \emph{[...]}   & $\bigcirc$ & $\bigcirc$ & $\bigcirc$ \\
        \emph{[App n]} & $\bigcirc$ & $\bigcirc$ & $\bigcirc$ \\
        \bottomrule
    \end{tabular*}

\clearpage

    \item Which of these is a sport?
    \label{appendix:main-survey:Q6}
    \begin{multicols}{4}
        \begin{answers}
            \item Hamburger
            \item Basketball
            \item Bathroom
            \item Skyscraper
        \end{answers}
    \end{multicols}

\emph{[Questions \ref{appendix:main-survey:Q7} through \ref{appendix:main-survey:Q14} are asked repeatedly for the newest app, oldest app, and a random app.]}

\begin{center}
\emph{Participants view a selected app from their third-party apps.}
\end{center}

    \item Do you recall authorizing \emph{[App name]}?
    \label{appendix:main-survey:Q7}
    \begin{multicols}{3}
        \begin{answers}
            \item Yes
            \item No
            \item Unsure
        \end{answers}
    \end{multicols}
    
    \item When was the last time you recall using \emph{[App name]}?
    \label{appendix:main-survey:Q8}
    \begin{multicols}{2}
        \begin{answers}
            \item Today
            \item In the previous week
            \item In the previous month
            \item In the previous year
            \item More than a year ago
            \item Unsure
        \end{answers}
    \end{multicols}

    \item Prior to seeing the details about \emph{[App name]}, were you aware this app had permission to access parts of your Google account data?
    \label{appendix:main-survey:Q9}
    \begin{multicols}{3}
        \begin{answers}
            \item Yes
            \item No
            \item Unsure
        \end{answers}
    \end{multicols}

    \item Please indicate how strongly you agree or disagree with the following:
    \label{appendix:main-survey:Q10}
    \begin{tiny}
    \hspace*{\fill}Strongly disagree | Disagree | Neither agree nor disagree | Agree | Strongly agree \\
    \end{tiny}
    \begin{scriptsize}
    \begin{tabularx}{\linewidth}{@{}l@{}C{1}@{}C{1}@{}C{1}@{}C{1}@{}C{1}@{}} 
        \toprule
        It's beneficial to me for \emph{[App name]} to\\ have access to my Google account.& $\bigcirc$ & $\bigcirc$ & $\bigcirc$ & $\bigcirc$ & $\bigcirc$ \\
        I'm concerned with \emph{[App name]} having\\ access to my Google account.& $\bigcirc$ & $\bigcirc$ & $\bigcirc$ & $\bigcirc$ & $\bigcirc$ \\
        I want to change which parts of my Google account\\ that \emph{[App name]} can access.& $\bigcirc$ & $\bigcirc$ & $\bigcirc$ & $\bigcirc$ & $\bigcirc$ \\
        \bottomrule
    \end{tabularx}
    \end{scriptsize}

    \item \emph{[App name]} holds the following permissions to access parts of your Google account. How **confident** are you that you understand what each permission allows the app to do?
    \label{appendix:main-survey:Q11}
    
    \begin{tiny}
    \hspace*{\fill}Not confident | Slightly confident | Moderately confident | Confident | Very confident \\
    \end{tiny}
    \begin{scriptsize}
    \begin{tabularx}{\linewidth}{@{}l@{}C{1}@{}C{1}@{}C{1}@{}C{1}@{}C{1}@{}}
        \toprule
        \emph{[App Permission 1]} & $\bigcirc$ & $\bigcirc$ & $\bigcirc$ & $\bigcirc$ & $\bigcirc$ \\
        \emph{[App Permission 2]} & $\bigcirc$ & $\bigcirc$ & $\bigcirc$ & $\bigcirc$ & $\bigcirc$ \\
        \emph{[...]} & $\bigcirc$ & $\bigcirc$ & $\bigcirc$ & $\bigcirc$ & $\bigcirc$ \\
        \emph{[App Permission n]}& $\bigcirc$ & $\bigcirc$ & $\bigcirc$ & $\bigcirc$ & $\bigcirc$ \\
        \bottomrule
    \end{tabularx}
    \end{scriptsize}

    \item \emph{[App name]} holds the following permissions to access parts of your Google account. How **necessary** do you think each permission is for the app to function in a way that benefits you?
    \label{appendix:main-survey:Q12}
    
    \begin{tiny}
    \hspace*{\fill}Not necessary | Slightly necessary | Moderately necessary | Necessary | Very necessary \\
    \end{tiny}
    \begin{scriptsize}
    \begin{tabularx}{\linewidth}{@{}l@{}C{1}@{}C{1}@{}C{1}@{}C{1}@{}C{1}@{}}
        \toprule
        \emph{[App Permission 1]} & $\bigcirc$ & $\bigcirc$ & $\bigcirc$ & $\bigcirc$ & $\bigcirc$ \\
        \emph{[App Permission 2]} & $\bigcirc$ & $\bigcirc$ & $\bigcirc$ & $\bigcirc$ & $\bigcirc$ \\
        \emph{[...]} & $\bigcirc$ & $\bigcirc$ & $\bigcirc$ & $\bigcirc$ & $\bigcirc$ \\
        \emph{[App Permission n]}& $\bigcirc$ & $\bigcirc$ & $\bigcirc$ & $\bigcirc$ & $\bigcirc$ \\
        \bottomrule
    \end{tabularx}
    \end{scriptsize}

    \item \emph{[App name]} holds the following permissions to access parts of your Google account. How **concerned** are you about the app accessing your account using these permissions?
    \label{appendix:main-survey:Q13}
    
    \begin{tiny}
    \hspace*{\fill}Not concerned | Slightly concerned | Moderately concerned | Concerned | Very concerned \\
    \end{tiny}
    \begin{scriptsize}
    \begin{tabularx}{\linewidth}{@{}l@{}C{1}@{}C{1}@{}C{1}@{}C{1}@{}C{1}@{}}
        \toprule
        \emph{[App Permission 1]} & $\bigcirc$ & $\bigcirc$ & $\bigcirc$ & $\bigcirc$ & $\bigcirc$ \\
        \emph{[App Permission 2]} & $\bigcirc$ & $\bigcirc$ & $\bigcirc$ & $\bigcirc$ & $\bigcirc$ \\
        \emph{[...]} & $\bigcirc$ & $\bigcirc$ & $\bigcirc$ & $\bigcirc$ & $\bigcirc$ \\
        \emph{[App Permission n]}& $\bigcirc$ & $\bigcirc$ & $\bigcirc$ & $\bigcirc$ & $\bigcirc$ \\
        \bottomrule
    \end{tabularx}
    \end{scriptsize}

    \item Please describe any concerns you have about \emph{[App name]} holding these permissions.
    \label{appendix:main-survey:Q14} (short answer)


\textbf{Manage third-party apps \& services with access to your Google Account}
\begin{itemize}[nosep]
    \item By now, you have seen information about apps and websites drawn from the ``Apps with access to your account'' page on your Google account (similar to the example below). 
    \item You can refer to that page when answering the following questions.
\end{itemize}

\textbf{Third-party apps with account access}
\begin{itemize}[nosep]
    \item You gave these sites and apps access to some of your Google Account data, including info that may be sensitive.  
    \item Remove access for those you no longer trust or use.
\end{itemize}

\textbf{Signing in with Google}
\begin{itemize}[nosep]
    \item You use your Google Account to sign in to these sites and apps. 
    \item They can view your name, email address, and profile picture. 
\end{itemize}
 
    \item The ``Apps with access to your account'' page helps me to better understand which third-party apps and websites are linked to my Google account.
    \label{appendix:main-survey:Q15}
    \begin{multicols}{2}
        \begin{answers}
            \item Strongly disagree
            \item Disagree
            \item Neither agree nor disagree
            \item Agree
            \item Strongly agree
        \end{answers}
    \end{multicols}

    \item The ``Apps with access to your account'' page helps me to better understand what parts of my Google account third-party apps can access.
    \label{appendix:main-survey:Q16}
    \begin{multicols}{2}
        \begin{answers}
            \item Strongly disagree
            \item Disagree
            \item Neither agree nor disagree
            \item Agree
            \item Strongly agree
        \end{answers}
    \end{multicols}

    \item After completing this survey, do you see yourself changing any settings on your ``Apps with access to your account'' page?
    \label{appendix:main-survey:Q17}
    \begin{multicols}{3}
        \begin{answers}
            \item Yes
            \item No
            \item Unsure
        \end{answers}
    \end{multicols}


    \item In six months do you see yourself reviewing third-party apps from your ``Apps with access to your account'' page?
    \label{appendix:main-survey:Q18}
    \begin{multicols}{3}
        \begin{answers}
            \item Yes
            \item No
            \item Unsure
        \end{answers}
    \end{multicols}

    \item You have indicated that you would change settings on your ``Apps with access to your account'' page. Please describe which settings would you change. \emph{[Shown if \ref{appendix:main-survey:Q17} is \enquote{Yes}]}
    \label{appendix:main-survey:Q19} (short answer)

    \item What would you look for when you review the ``Apps with access to your account'' page in six months? 
    \emph{[Shown if \ref{appendix:main-survey:Q18} is \enquote{Yes}]}
    \label{appendix:main-survey:Q20} (short answer)
 
    \item What new features (if any) would you like to add to the ``Apps with access to your account'' page?
    \label{appendix:main-survey:Q21} (short answer)
    
    \item Suppose Google sent an email reminder to review your ``Apps with access to your account'' page. How often would you like to be reminded?
    \label{appendix:main-survey:Q22}
    \begin{multicols}{2}
        \begin{answers}
            \item Once a week
            \item Once a month
            \item Once a year
            \item Once every three years
            \item I do not want to be reminded
        \end{answers}
    \end{multicols}

    \item Suppose Google required you to reapprove the third-party apps on your Google account. How often would you like to reapprove apps?
    \label{appendix:main-survey:Q23}
    \begin{multicols}{2}
        \begin{answers}
            \item Once a week
            \item Once a month
            \item Once a year
            \item Once every three years
            \item I do not want to reapprove apps
        \end{answers}
    \end{multicols}

    \item Which of these is cold?
    \label{appendix:main-survey:Q24}
    \begin{multicols}{4}
        \begin{answers}
            \item Fire
            \item Summer
            \item Lava
            \item Ice cream
        \end{answers}
    \end{multicols}

    \item Rather than approve all permissions when installing third-party apps, I would want third-party apps to seek my approval each time they access my Google account.
    \label{appendix:main-survey:Q25}
    \begin{multicols}{2}
        \begin{answers}
            \item Strongly disagree
            \item Disagree
            \item Neither agree nor disagree
            \item Agree
            \item Strongly agree
        \end{answers}
    \end{multicols}

    \item I would want to designate specific data (eg. certain emails, individual contacts, particular calendar events) as private and inaccessible to third-party apps.
    \label{appendix:main-survey:Q26}
    \begin{multicols}{2}
        \begin{answers}
            \item Strongly disagree
            \item Disagree
            \item Neither agree nor disagree
            \item Agree
            \item Strongly agree
        \end{answers}
    \end{multicols}

\end{questions}

\end{footnotesize}
\clearpage

\onecolumn
\section{Additional Figures and Tables}\label{appendix:additional}

\begin{table*}[h]
    \caption{The top 26 (4 apps tied for 23rd ranked by authorized count) apps with access, the categories for the requested permissions, the average number of days authorized to the participants' Google account, and the number of permissions each app requested.\label{tab:apps}}
    \footnotesize \centering
    \begin{tabular}{r|l|c|c|c}
    
                                                    &                                   & \textbf{Number}  & \textbf{Avg Num of}       & \textbf{Number of}    \\
        \textbf{App}                           & \textbf{Permission Categories}    &\textbf{Authorized}     & \textbf{Days Authorized}   & \textbf{Permissions}  \\
        \toprule
        Dropbox                                     & Basic account info, Google Contacts & 40                & 502                       & 4                     \\
        SAMSUNG Account                             & Additional access, Basic account info & 38                & 63                        & 5                     \\
        Shop: package \& order tracker              & Basic account info, Gmail         & 32                & 37                        & 3                     \\
        Nest                                        & Additional access, Basic account info & 31                & 265                       & 3                     \\
        Windows                                     & Basic account info, Gmail, Google Calendar & 29                & 397                       & 8                     \\
                                                    & Google Contacts                   &                   &                           &                       \\
        Microsoft SwiftKey Keyboard                 & Additional access, Basic account info & 27                & 102                       & 3                     \\
        macOS                                       & Basic account info, Gmail, Google Calendar, & 19                & 452                       & 6                     \\
                                                    & Google Contacts, Google Hangouts    &                   &                           &                       \\
        Zoom                                        & Basic account info, Google Calendar,  & 16                & 158                       & 5         \\
                                                    & Google Contacts                       &                   &                           &           \\
        WhatsApp Messenger                          & Google Drive                      & 13                & 663                       & 2                     \\
        Rakuten Cash Back                           & Basic account info, Gmail         & 13                & 268                       & 3                     \\
        Pokémon GO                                  & Additional access, Basic account info & 10                & 307                       & 4                     \\
        Microsoft apps \& services                  & Additional access, Basic account info, Gmail, & 10            & 290                       & 8                     \\
                                                    & Google Calendar, Google Contacts, Google Drive &                       &                       \\
        Samsung Email                               & Basic account info, Gmail         & 9                 & 1,014                    & 4                     \\
        Paribus                                     & Basic account info, Gmail         & 9                 & 390                       & 8                     \\
        Unroll.me                                   & Basic account info, Gmail, Google Contacts & 8                 & 1,389                    & 5                     \\
        Clash Royale                                & Google Play                       & 8                 & 644                       & 1                     \\
        Google Nest Hub                             & Basic account info                & 8                 & 287                       & 1                     \\
        PlayStation Network                         & Basic account info, YouTube       & 7                 & 573                       & 5                     \\
        Shop                                        & Basic account info, Gmail         & 7                 & 441                       & 5                     \\
        Chromecast                                  & Basic account info                & 7                 & 225                       & 1                     \\
        Lumin PDF                                   & Basic account info, Google Drive  & 7                 & 164                       & 5                     \\
        DocHub - PDF Sign \& Edit                   & Basic account info, Gmail, Google Contacts & 7                 & 86                        & 6                     \\
                                                    & Google Drive                      &                   &                           &                       \\
        IFTTT                                       & Basic account info                & 6                 & 1,130                    & 1                     \\
        Sweatcoin — Walking step counter \& tracker & Basic account info                & 6                 & 486                       & 2                     \\
        YouTube on Xbox Live                        & YouTube                           & 6                 & 420                       & 2                     \\
        Fetch Rewards                               & Basic account info, Gmail         & 6                 & 159                       & 3                     \\
    \end{tabular}
\end{table*}

\clearpage


\begin{table*}[h]
    \caption{The top 25 authorized SSO ranked by the number of authorized accesses to participants' Google accounts, and the number of permissions each SSO requested.\label{tab:ssos}}
    \footnotesize \centering
    \begin{tabular}{r|l|c|c}
    
                                                            &                                               & \textbf{Number}  & \textbf{Number of}    \\
        \textbf{SSO}                                        & \textbf{Permission Categories}                & \textbf{Authorized}    & \textbf{Permissions}  \\
        \toprule
        Movies Anywhere                                     & Basic account info, Google Play               & 167               & 3                     \\
        Honey                                               & Basic account info                            & 80                & 2                     \\
        Amazon Alexa                                        & Basic account info, Gmail, Google Calendar    & 72                & 4                     \\
        Quora                                               & Basic account info, Google Contacts           & 71                & 4                     \\
        Adobe                                               & Basic account info                            & 63                & 2                     \\
        Reddit                                              & Basic account info                            & 61                & 3                     \\
        Microsoft apps \& services                          & Gmail, Google Calendar,                       & 58                & 8                     \\
                                                            & Google Contacts, Google Drive                 &                   &                       \\
        Pinterest                                           & Basic account info                            & 58                & 2                     \\
        Windows                                             & Basic account info, Gmail,                    & 53                & 8                     \\
                                                            & Google Calendar, Google Contacts              &                   &                       \\
        Glassdoor                                           & Additional access, Basic account info         & 52                & 3                     \\
        The New York Times                                  & Basic account info                            & 49                & 2                     \\
        Doordash                                            & Basic account info                            & 46                & 2                     \\
        Spotify                                             & Basic account info                            & 44                & 2                     \\
        macOS                                               & Basic account info, Gmail, Google Calendar,   & 43                & 6                     \\
                                                            & Google Contacts, Google Hangouts              &                   &                       \\
        Quizlet                                             & Basic account info                            & 43                & 2                     \\
        Dropbox                                             & Basic account info, Google Contacts           & 41                & 4                     \\
        SAMSUNG Account                                     & Additional access, Basic account info         & 38                & 5                     \\
        Zoom                                                & Basic account info, Google Calendar,          & 35                & 5                     \\
                                                            & Google Contacts                               &                   &                       \\
        Shop: package \& order tracker                      & Basic account info, Gmail                     & 32                & 3                     \\
        Nest                                                & Additional access, Basic account info         & 31                & 3                     \\
        paypal.com                                          & Basic account info                            & 31                & 2                     \\
        Adobe Acrobat Reader: PDF Viewer, Editor \& Creator & Basic account info, Google Drive              & 30                & 5                     \\        
        Best Buy App                                        & Basic account info                            & 30                & 2                     \\
        Shop                                                & Basic account info, Gmail                     & 30                & 3                     \\
        SoundCloud                                          & Basic account info                            & 29                & 2                     \\
    \end{tabular}
\end{table*}

\clearpage


\renewcommand{\arraystretch}{1.25}
\begin{table*}[h]
\caption{The top 28 (5-way tie for permissions count of 8) most requested permissions by third-party apps with authorized access to participants' Google accounts, the permission category, and a count of how many of each permissions were authorized.\label{tab:app-permissions}}
\footnotesize \centering
\begin{tabular}{r|p{2.3cm}|l}
     \textbf{Count} & \textbf{Category}     & \textbf{Permission} \\
     \toprule
     223            & Google Play           & Create, edit, and delete your Google Play Games activity \\
     189            & Basic account info    & See your primary Google Account email address \\
     177            & Basic account info    & See your personal info, including any personal info you've made publicly available \\
     71             & Google Drive          & See, create, and delete its own configuration data in your Google Drive \\
     44             & Basic account info    & Associate you with your personal info on Google \\
     28             & Google Drive          & See, edit, create, and delete only the specific Google Drive files you use with this app \\
     27             & Google Drive          & See, edit, create, and delete all of your Google Drive files \\
     27             & Gmail                 & Read, compose, send, and permanently delete all your email from Gmail \\
     26             & Google Contacts       & See, edit, download, and permanently delete your contacts \\
     24             & Google Calendar       & See, edit, share, and permanently delete all the calendars you can access using Google Calendar \\
     18             & Google Contacts       & See and download your contacts \\
     16             & Basic account info    & Full account access \\
     14             & Additional access     & See and download your exact date of birth \\
     13             & Additional access / Basic account info    & Display and run third-party web content in prompts and sidebars inside Google applications \\
     12             & Gmail                 & View your email messages and settings \\
     12             & Google Drive          & See and download all your Google Drive files \\
     11             & Additional access     & See your age group \\
     11             & Additional access / Basic account info     & Use Google Fit to see and store your physical activity data \\
     10             & Google Docs           & See, edit, create, and delete your spreadsheets in Google Drive \\
     10             & Google Drive          & Connect itself to your Google Drive \\
     10             & Additional access     & Connect to an external service \\
     9              & Additional access     & See your language preferences \\
     9              & Additional access / Basic account info   & See and add to your Google Fit physical activity data \\
     8              & Additional access / Basic account info   & See and add info about your body measurements and heart rate to Google Fit \\
     8              & Gmail                 & Send email on your behalf\\
     8              & Google Docs           & See, create, and edit all Google Docs documents you have access to\\
     8              & Additional access     & See and download all of your Google Account email addresses\\
     8              & YouTube               & Manage your YouTube account\\
\end{tabular}
\end{table*}

\clearpage


\begin{table*}[h]
\caption{The top 25 most requested permissions by SSO with authorized access to participants' Google accounts ranked by the number of each permission authorized, and the permission category.\label{tab:sso-permissions}}
\footnotesize \centering
\begin{tabular}{r|l|l}
     \textbf{Count} & \textbf{Category} & \textbf{Permission} \\
     \toprule
     976            & Basic account info    & See your primary Google Account email address \\
     938            & Basic account info    & See your personal info, including any personal info you've made publicly available \\
     88             & Basic account info    & Associate you with your personal info on Google \\
     45             & Google Play           & Create, edit, and delete your Google Play Games activity \\
     27             & Google Drive          & See, edit, create, and delete only the specific Google Drive files you use with this app \\
     25             & Gmail                 & Read, compose, send, and permanently delete all your email from Gmail \\
     25             & Google Calendar       & See, edit, share, and permanently delete all the calendars you can access using Google Calendar \\
     24             & Google Contacts       & See, edit, download, and permanently delete your contacts \\
     20             & Google Contacts       & See and download your contacts \\
     20             & Google Drive          & See, edit, create, and delete all of your Google Drive files \\
     16             & Google Drive          & See, create, and delete its own configuration data in your Google Drive \\
     14             & Additional access     & See and download your exact date of birth \\
     12             & Gmail                 & View your email messages and settings \\
     11             & Additional access     & See your age group \\
     11             & Google Drive          & See and download all your Google Drive files \\
     10             & Google Docs           & See, edit, create, and delete your spreadsheets in Google Drive \\
     10             & Google Drive          & Connect itself to your Google Drive \\
     10             & Additional access     & Display and run third-party web content in prompts and sidebars inside Google applications \\
     9              & Additional access     & See your language preferences \\
     9              & Gmail                 & Send email on your behalf \\
     9              & Additional access     & See and add info about your body measurements and heart rate to Google Fit \\
     9              & Additional access     & See and add to your Google Fit physical activity data \\
     9              & Additional access     & Connect to an external service \\
     8              & Additional access     & See and download all of your Google Account email addresses \\
     8              & Additional access     & Use Google Fit to see and store your physical activity data \\
\end{tabular}
\end{table*}

\clearpage


\begin{table}[ht]
    \caption{Full demographics data of the participants of the first survey. \label{tab:demographicsFull}}
    \centering
    \begin{tabular}{r|rrrrrrrr|rr}
    \hline
    & \multicolumn{2}{c}{\textbf{Male}} & \multicolumn{2}{c}{\textbf{Female}}
    & \multicolumn{2}{c}{\textbf{Non-binary}} & \multicolumn{2}{c|}{\textbf{PND}} & \multicolumn{2}{c}{\textbf{Total}} \\
    & \textbf{No.} & \textbf{\%} & \textbf{No.} & \textbf{\%}
    & \textbf{No.} & \textbf{\%} & \textbf{No.} & \textbf{\%} & \textbf{No.} & \textbf{\%}\\
    \hline
    \textbf{Age} & 216 & 50 & 205 & 47 & 11 & 3 & 2 & 0 & 434 & 100\\
    \hline
    18 - 24 & 58 & 13 & 71 & 16 & 4 & 1 & 0 & 0 & 133 & 31\\
    25 - 34 & 84 & 19 & 67 & 15 & 6 & 1 & 0 & 0 & 157 & 36\\
    35 - 44 & 44 & 10 & 30 & 7 & 1 & 0 & 0 & 0 & 75 & 17\\
    45 - 54 & 19 & 4 & 25 & 6 & 0 & 0 & 0 & 0 & 44 & 10\\
    55 - 64 & 7 & 2 & 8 & 2 & 0 & 0 & 0 & 0 & 15 & 3\\
    65 or older & 4 & 1 & 4 & 1 & 0 & 0 & 0 & 0 & 8 & 2\\
    Prefer not to disclose & 0 & 0 & 0 & 0 & 0 & 0 & 2 & 0 & 2 & 0\\
    \hline
    \textbf{Highest level of school} & 216 & 50 & 205 & 47 & 11 & 3 & 2 & 0 & 434 & 100\\
    \hline
    No schooling completed & 0 & 0 & 0 & 0 & 0 & 0 & 0 & 0 & 0 & 0\\
    Some high school & 3 & 1 & 1 & 0 & 0 & 0 & 0 & 0 & 4 & 1\\
    High school & 20 & 5 & 17 & 4 & 2 & 0 & 0 & 0 & 39 & 9\\
    Some college & 48 & 11 & 37 & 9 & 4 & 1 & 0 & 0 & 89 & 21\\
    Trade & 4 & 1 & 5 & 1 & 0 & 0 & 0 & 0 & 9 & 2\\
    Associate's Degree & 19 & 4 & 7 & 2 & 1 & 0 & 0 & 0 & 27 & 6\\
    Bachelor's Degree & 89 & 21 & 91 & 21 & 3 & 1 & 0 & 0 & 183 & 42\\
    Master's Degree & 24 & 6 & 36 & 8 & 1 & 0 & 0 & 0 & 61 & 14\\
    Professional degree & 4 & 1 & 4 & 1 & 0 & 0 & 0 & 0 & 8 & 2\\
    Doctorate & 5 & 1 & 6 & 1 & 0 & 0 & 0 & 0 & 11 & 3\\
    Prefer not to disclose & 0 & 0 & 1 & 0 & 0 & 0 & 2 & 0 & 3 & 1\\
    \hline
    \textbf{Background} & 216 & 50 & 205 & 47 & 11 & 3 & 2 & 0 & 434 & 100\\
    \hline
    Technical & 74 & 17 & 25 & 6 & 1 & 0 & 0 & 0 & 100 & 23\\
    Non-Technical & 134 & 31 & 178 & 41 & 10 & 2 & 0 & 0 & 322 & 74\\
    Prefer not to disclose & 8 & 2 & 2 & 0 & 0 & 0 & 2 & 0 & 12 & 3\\
    \end{tabular}
\end{table}

\clearpage

\begin{table}[ht]
    \caption{Full demographics data of the participants of the second survey. \label{tab:demographicsMainStudy}}
    \centering
    \begin{tabular}{r|rrrrrrrr|rr}
    \hline
    & \multicolumn{2}{c}{\textbf{Male}} & \multicolumn{2}{c}{\textbf{Female}} 
    & \multicolumn{2}{c}{\textbf{Non-binary}} & \multicolumn{2}{c|}{\textbf{PND}} & \multicolumn{2}{c}{\textbf{Total}} \\
    & \textbf{No.} & \textbf{\%} & \textbf{No.} & \textbf{\%}
    & \textbf{No.} & \textbf{\%} & \textbf{No.} & \textbf{\%} & \textbf{No.} & \textbf{\%}\\
    \hline
    \textbf{Age} & 68 & 46 & 75 & 51 & 4 & 3 & 1 & 1 & 148 & 100\\
    \hline
    18 - 24 & 18 & 12 & 34 & 23 & 3 & 2 & 0 & 0 & 55 & 37\\
    25 - 34 & 24 & 16 & 21 & 14 & 1 & 1 & 0 & 0 & 46 & 31\\
    35 - 44 & 15 & 10 & 9 & 6 & 0 & 0 & 0 & 0 & 24 & 16\\
    45 - 54 & 7 & 5 & 7 & 5 & 0 & 0 & 0 & 0 & 14 & 9\\
    55 - 64 & 2 & 1 & 3 & 2 & 0 & 0 & 0 & 0 & 5 & 3\\
    65 or older & 2 & 1 & 1 & 1 & 0 & 0 & 0 & 0 & 3 & 2\\
    Prefer not to disclose & 0 & 0 & 0 & 0 & 0 & 0 & 1 & 1 & 1 & 1\\
    \hline
    \textbf{Highest level of school} & 68 & 46 & 75 & 51 & 4 & 3 & 1 & 1 & 148 & 100\\
    \hline
    No schooling completed & 0 & 0 & 0 & 0 & 0 & 0 & 0 & 0 & 0 & 0\\
    Some high school & 0 & 0 & 0 & 0 & 0 & 0 & 0 & 0 & 0 & 0\\
    High school & 7 & 5 & 3 & 2 & 1 & 1 & 0 & 0 & 11 & 7\\
    Some college & 20 & 14 & 13 & 9 & 2 & 1 & 0 & 0 & 35 & 24\\
    Trade & 1 & 1 & 2 & 1 & 0 & 0 & 0 & 0 & 3 & 2\\
    Associate's Degree & 2 & 1 & 1 & 1 & 0 & 0 & 0 & 0 & 3 & 2\\
    Bachelor's Degree & 28 & 19 & 40 & 27 & 1 & 1 & 0 & 0 & 69 & 47\\
    Master's Degree & 8 & 5 & 13 & 9 & 0 & 0 & 0 & 0 & 21 & 14\\
    Professional degree & 1 & 1 & 1 & 1 & 0 & 0 & 0 & 0 & 2 & 1\\
    Doctorate & 1 & 1 & 1 & 1 & 0 & 0 & 0 & 0 & 2 & 1\\
    Prefer not to disclose & 0 & 0 & 1 & 1 & 0 & 0 & 1 & 1 & 2 & 1\\
    \hline
    \textbf{Background} & 68 & 46 & 75 & 51 & 4 & 3 & 1 & 1 & 148 & 100\\
    \hline
    Technical & 21 & 14 & 11 & 7 & 0 & 0 & 0 & 0 & 32 & 22\\
    Non-Technical & 44 & 30 & 64 & 43 & 4 & 3 & 0 & 0 & 112 & 76\\
    Prefer not to disclose & 3 & 2 & 0 & 0 & 0 & 0 & 1 & 1 & 4 & 3\\
    \end{tabular}
\end{table}

\clearpage

\begin{figure*}[ht]
  \begin{subfigure}{0.5\textwidth}
    \includegraphics[width=0.9\linewidth]{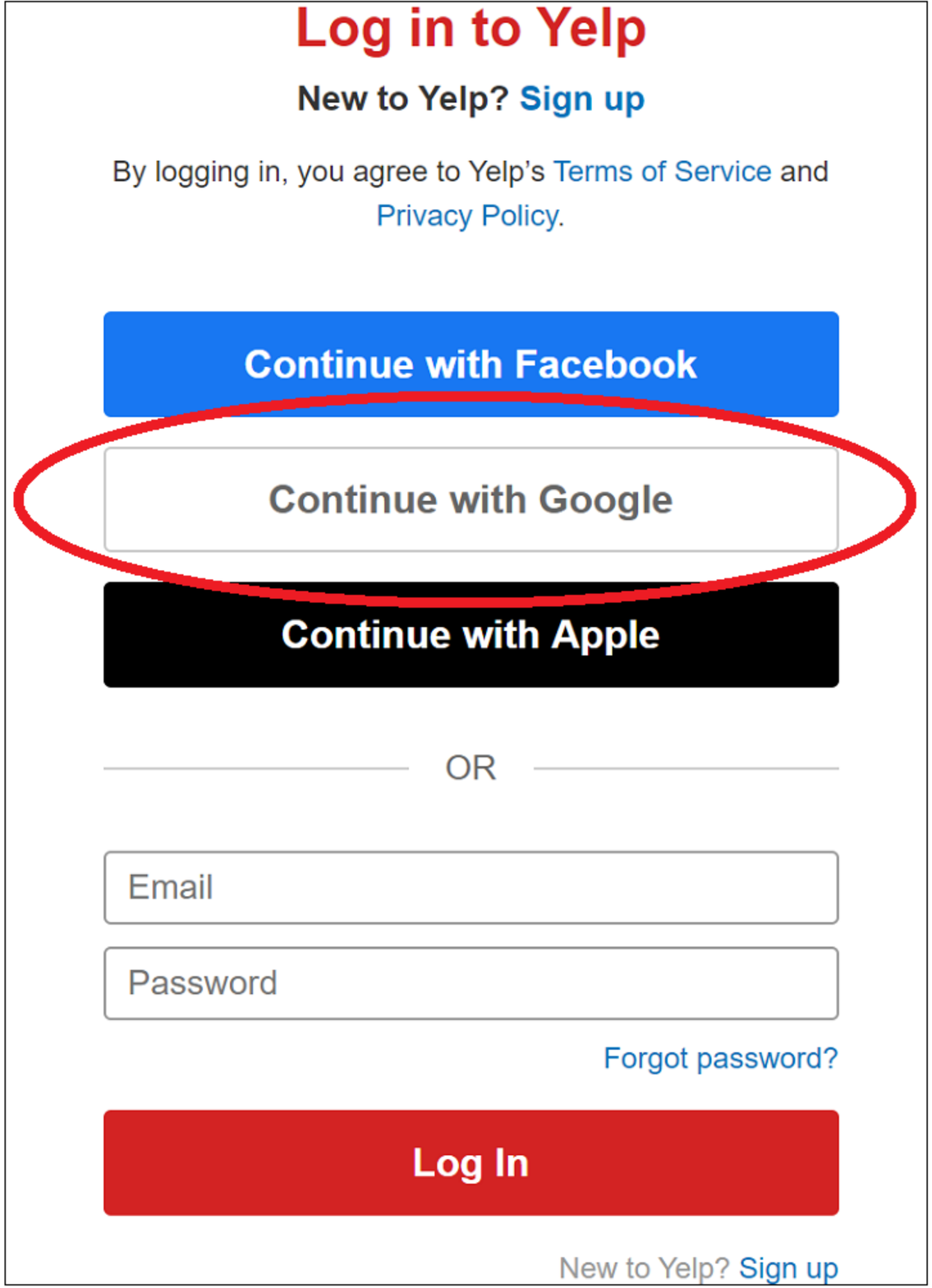} 
    \caption{Google account SSO prompt.}
    \label{fig:GoogleSSOContext}
  \end{subfigure}
  \begin{subfigure}{0.5\textwidth}
    \includegraphics[width=0.9\linewidth]{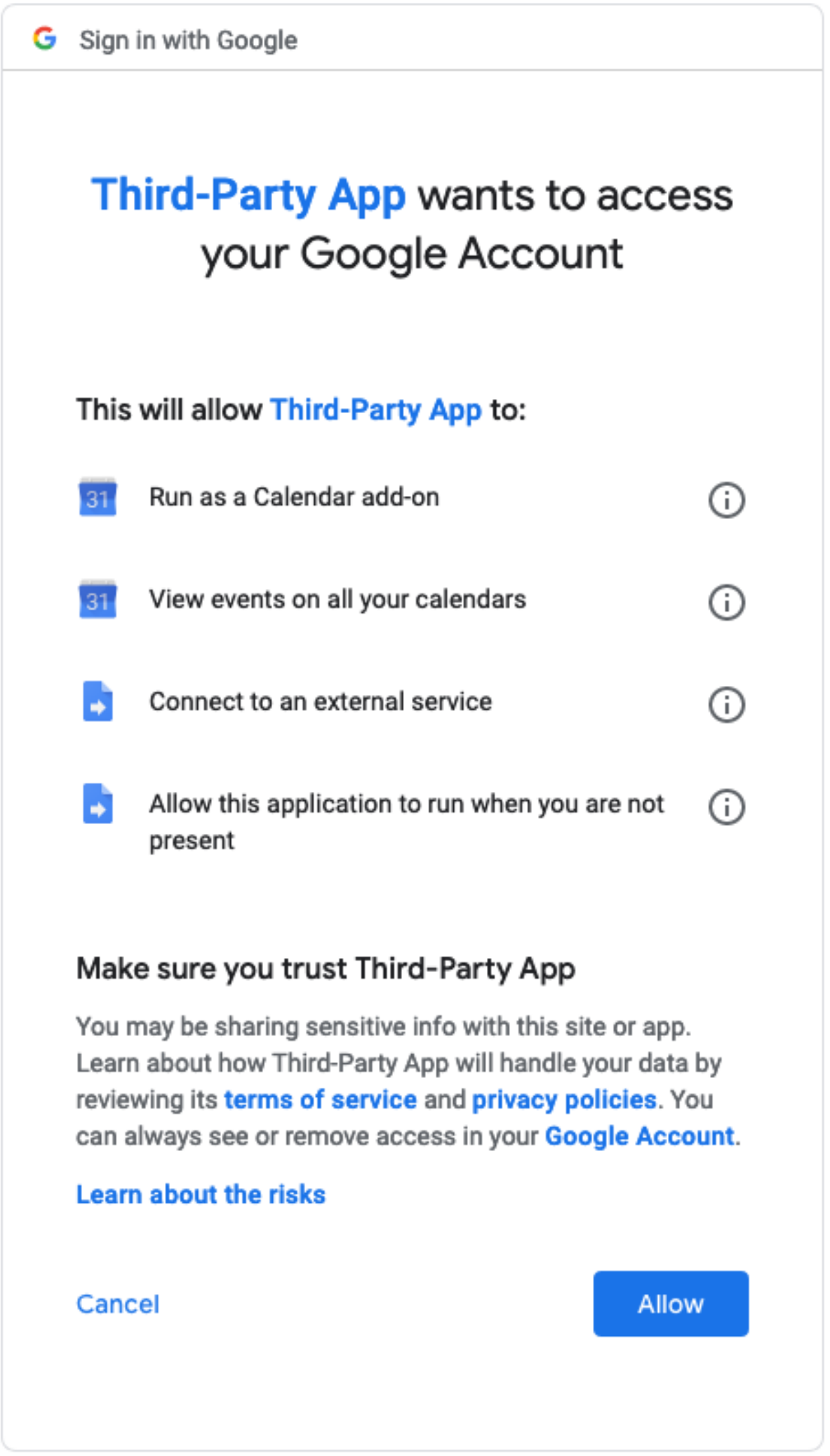}
    \caption{Third-party app access consent dialog.}
    \label{fig:GoogleThirdPartyContext}
  \end{subfigure}
  \caption{Google account access authorizations.}
  \label{fig:GoogleSSOContextAndGoolgeThirdPartyContext}
\end{figure*}

\clearpage

\begin{figure*}[ht]
    \centering
    \includegraphics[width=0.9\linewidth]{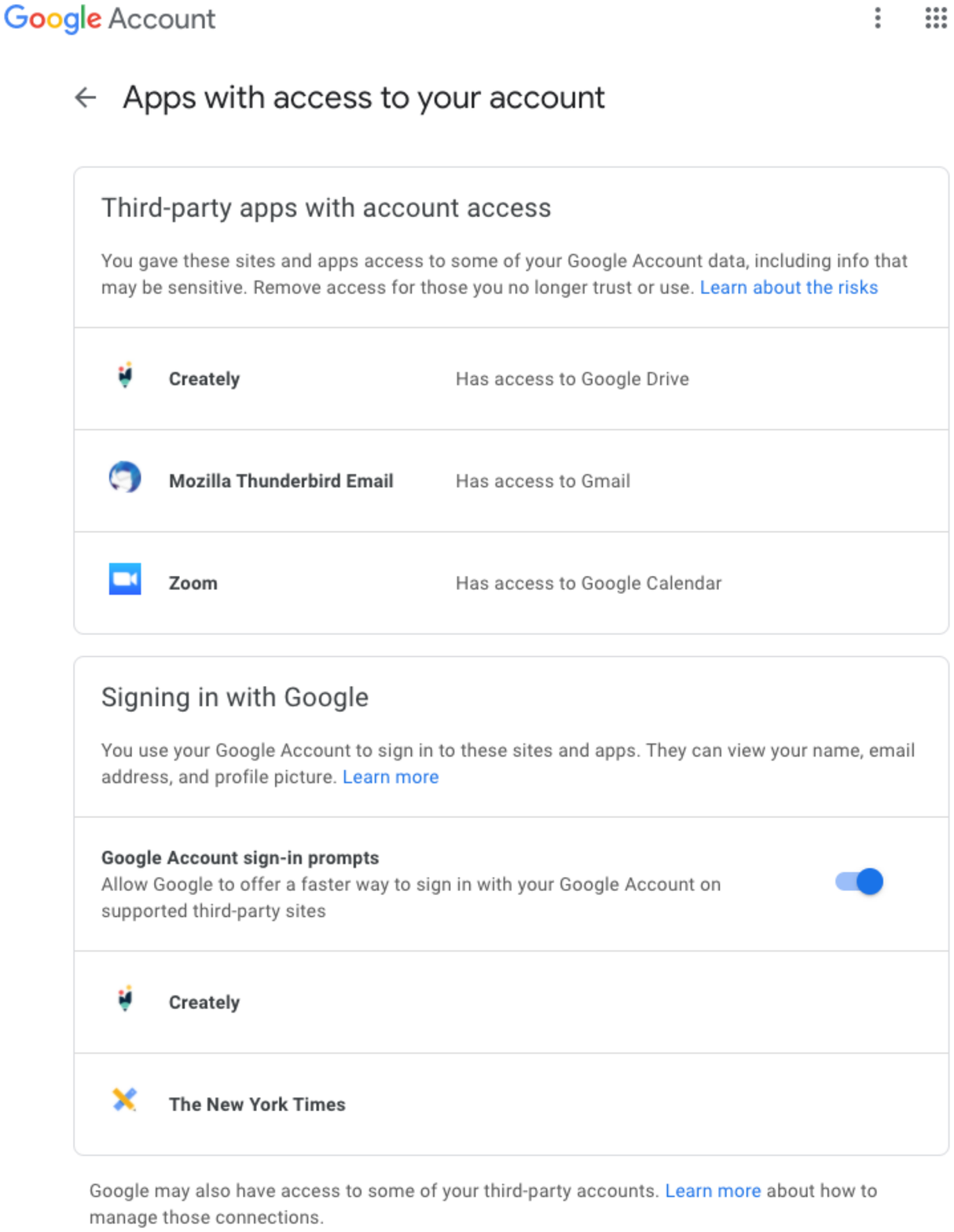} 
    \caption{Google's \enquote{Apps with access to your account} page.}
    \label{fig:AppsWithAccessToYourAccountImage}
\end{figure*}

\clearpage

\begin{figure*}[ht]
    \centering
    \includegraphics[width=0.9\linewidth]{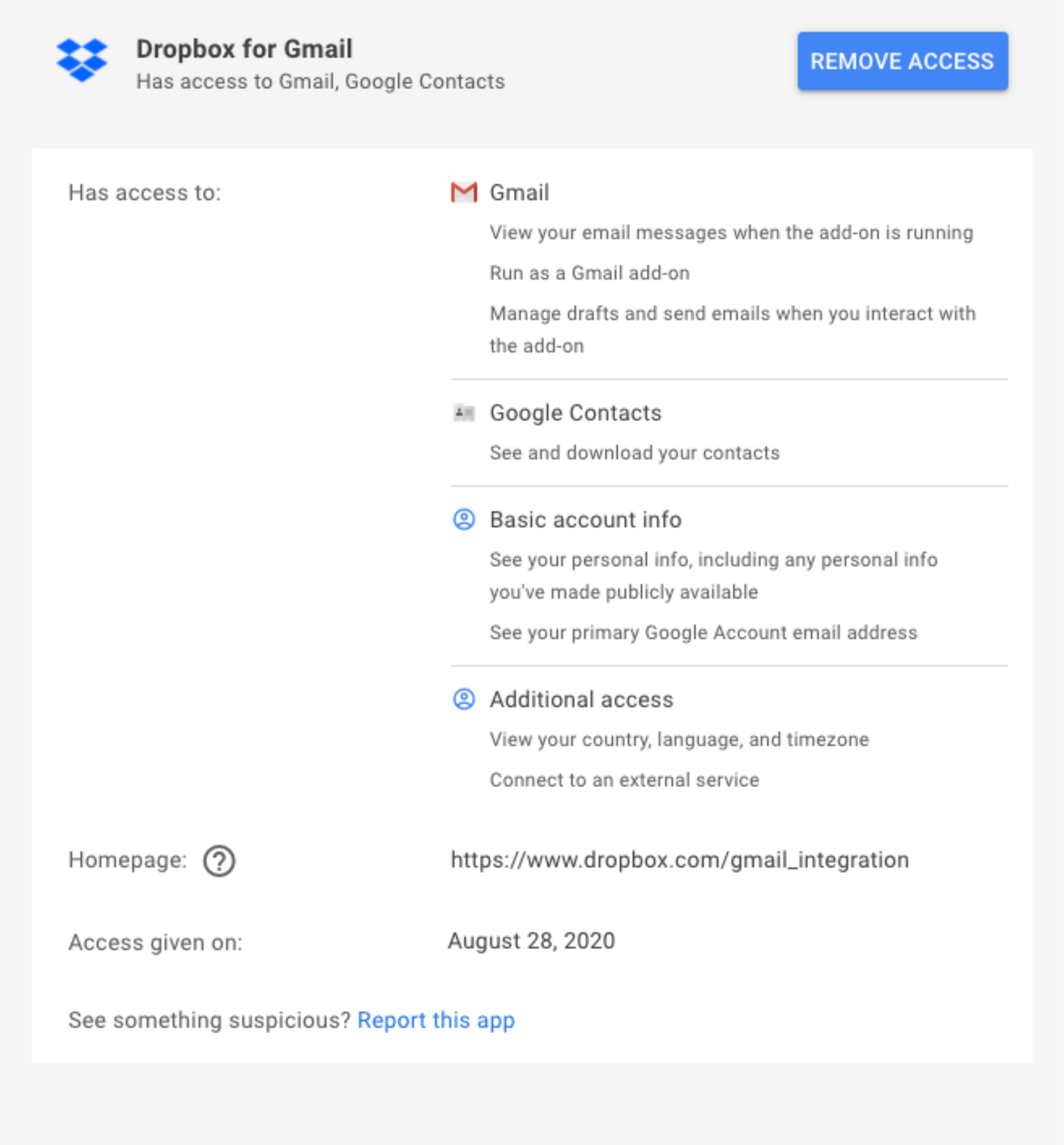}
    \caption{Dropbox for Gmail third-party app as displayed on Google's \enquote{Apps with access to your account} page.}
    \label{fig:DropboxAccessToYourAccountImage}
\end{figure*}

\clearpage

\clearpage

\twocolumn
\section{Qualitative Codebook}\label{appendix:codebook}
\begin{itemize}

\item\textbf{unconcerned (203)}

\emph{permissions-necessary (18), low-permission-level (6), use-often (2), trust-google (1), assume-secure (1), other-primary-email (1)}

\item\textbf{concerned (169)}

\emph{permissions (77), personal-data (59), unnecessary-access-to-data (39), misuse (18), data-leak (7), permissions-after-deleted-app (6), privacy (4), tracking (3), permissions-more-permissions-than-described (1), failure-to-update (1), access-when-not-using-app (1)}

\item\textbf{review-app-access (140)}

\emph{unused-apps (44), new-apps-with-access (22), what-data-is-accessible (16), do-not-remember-authorizing (12), permissions (10), how-much-access-allowed (8), accidentally-added (7), necessary-permissions-only (7), unfamiliar-apps (6), account-login (4), privacy (4), permissions-changed (4), how-long-access (3), suspicious-apps (3), unauthorized-apps (2), most-used (2), all (1), unauthorized-permissions (1), unnecessary-access (1), specific-app (1), full-account-access (1)}

\item\textbf{remove-app-access (99)}

\emph{unused-apps (57), some-apps (11), specific-app (9), unfamiliar-apps (7), all (6), apps-with-account-access (4), apps-with-google-drive-access (2)}

\item\textbf{security (96)}

\emph{information (9), misuse (7), information-leak (6), information-theft (3), website (1), app-owner (1), sign-in (1)}

\item\textbf{ease-of-use (94)}

\emph{account-creation (13), access-removal (5), sign-in-process (4), fills-in-information (2), how-long (1), already-connected (1), simple (1)}

\item\textbf{unknowns (92)}

\emph{information-access (35), use-of-information (15), how-data-used (10), no-unwanted-email (7), what-permission-allows (6), do-they-keep-information (3), email-access (3), why-access-needed (3), why-permissions-necessary (2), why-information-needed (2), who-can-access-data (2), associated-with-personal-info (2), payments (1), can-i-delete-my-data (1), google-guideline-enforcement (1), will-it-notify-on-data-delete (1), how-long-data-stored (1)}

\item\textbf{personal-information (89)}

\emph{what-type-accessible-collected (21), how-used (15), amount-collected (9), sensitive (8), is-shared (5), sharing (4), account (3), email (3), calendar (1), kept-confidential (1), photos (1), schedule (1), account-password (1), accounts (1)}

\item\textbf{none (86)}

\item\textbf{permissions (85)}

\emph{necessary (26), access (20), what-information-accessible (13), email-access (5), correct (3), contacts (3), calendar (2), modify-data (2), media (2), location (2), contact-info (1), local-files (1), google-drive (1), webcam (1), alternate-app-with-fewer (1), browsing-history (1)}

\item\textbf{privacy (76)}

\emph{data-protection (2), data-used-for-advertising (1), privacy-policy (1)}

\item\textbf{more-transparency (61)}

\emph{app-usage-log (13), detailed-permission-explanation (12), when-access-authorized (8), misuse-reports (6), what-data-accessible (5), data-access-logs (5), what-app-needs-to-work (3), what-parts-of-account-accessible (3), which-app-accessed-the-most (1), information-about-security (1), how-data-used (1), company-providing-app (1), how-access-was-authorized (1), terms-and-conditions (1), location-of-app-access (1), listing-of-sites-indirectly-accessing (1), vulnerability-alert (1), permission-usage-log (1), information-about-privacy (1), how-to-remove-information (1)}

\item\textbf{is-trusted (56)}

\emph{app (26), website (13), company (5), developer (2)}

\item\textbf{utility (45)}

\emph{synch-info (13), file-transfer (6), scan-documents (3), email (3), screenshot (2), edit-pdf (2), install (1), webcam (1), zip-extractor (1), manage-apps (1)}

\item\textbf{other-options (43)}

\emph{facebook (14), apple (2), use-without-sign-in (1)}

\item\textbf{nothing (41)}

\item\textbf{calendar (38)}

\item\textbf{specific-app-or-service (37)}

\emph{zoom (7), google-drive (4), google-docs (2), youtube (1), only-office (1), calendly (1), grammarly (1), google-audio-player (1), snapchat (1), gleam (1), twitter (1), honey (1), slack (1), movies-anywhere (1), plex (1), paypal (1), Doodle (1), google-navigation (1), peloton (1), yelp (1), headspace (1), mcdonalds (1), nest (1), amazon (1), google-play (1), calendars-5 (1), todoist (1), linked-in (1), google-sheets (1)}

\item\textbf{trust (36)}

\emph{website (13), app (11), company (5), google (3), site (1), no-unwanted-email (1)}

\item\textbf{is-app-useful (31)}

\item\textbf{change-permissions (30)}

\emph{limit-access (8), remove-unnecessary (4), contacts (4), account-info (2), specific-app (2), email (2), unused-apps (2), calendar (1), delete-files (1), storage (1), sharing (1), google-drive (1), change-files (1)}

\item\textbf{tradeoff (30)}

\emph{useful (11), sharing-information (4), frequent-use (1), security (1), account-creation (1), length-of-use (1), frequent-use (1), another-account (1)}

\item\textbf{sign-in (27)}

\item\textbf{infrequent-use (27)}

\item\textbf{do-not-recall-authorizing (24)}

\emph{permission (6), app (4)}

\item\textbf{app-beneficial (22)}

\item\textbf{review-changes (22)}

\emph{changed-without-notification (2)}

\item\textbf{trust-app (22)}

\item\textbf{safety-of-app (21)}

\item\textbf{remember-login (20)}

\item\textbf{notifications (19)}

\emph{reminders (5), upon-changes (2), to-review (2), remove-access-after-app-removed (1), data-breach (1), when-account-accessed (1), email (1), unused-apps (1)}

\item\textbf{gaming (19)}

\item\textbf{contacts (18)}

\item\textbf{easier-access-removal (17)}

\item\textbf{improved-user-interface (17)}

\emph{easier-to-access-page (4), sort-apps-by-permission-type (3), personalization (1), order-by-date-authorized (1), edit-preferences-directly (1), color-code-based-on-level-of-access (1), add-app-description-text (1), specific-icons (1), deletion-suggestions (1), filter-by-authorization-date (1), opt-out-option (1)}

\item\textbf{no-recall (15)}

\item\textbf{personal-info (15)}

\emph{health-data (2), location (1), age (1)}

\item\textbf{necessary (15)}

\item\textbf{shopping (14)}

\item\textbf{wants-to-protect-personal-data (14)}

\emph{leaks (1)}

\item\textbf{unsure (14)}

\item\textbf{convenience (12)}

\emph{account-creation (4)}

\item\textbf{transfer-of-trust (11)}

\emph{from-google (10)}

\item\textbf{distrust (11)}

\emph{no-unwanted-email (5), google (5), app (1), company (1)}

\item\textbf{fewer-accounts (10)}

\item\textbf{permission-level-control (9)}

\emph{remove-individual-permissions (4), set-some-permissions-never-use (1)}

\item\textbf{unwanted-emails (7)}

\item\textbf{ability-to-limit-access (7)}

\emph{temporarily-block-app-access (2), minimum-necessary (1), remove-access-if-not-used (1), temporarily-allow-app-access (1), time-limits (1)}

\item\textbf{will-remove-app-access (6)}

\item\textbf{review-how-trusted-app-is (5)}

\item\textbf{photos (4)}

\item\textbf{comfort-with-app-or-service (4)}

\emph{access-to-information (4)}

\item\textbf{work-related (4)}

\item\textbf{trade-off-for-convenience (4)}

\item\textbf{ability-to-secure-data (3)}

\emph{limit-what-data-can-be-accessed (1)}

\item\textbf{require-reauthorization (3)}

\emph{yearly (1), period-of-time (1), unused-apps (1)}

\item\textbf{will-review-app-access (3)}

\item\textbf{uncertain (2)}

\item\textbf{use-separate-account-for-app-access (2)}

\item\textbf{assess-risks (2)}

\item\textbf{resigned (2)}

\item\textbf{app-not-beneficial (2)}

\item\textbf{education (1)}

\item\textbf{change-sso-access (1)}

\item\textbf{remove-app-accessreview-app-access (1)}

\item\textbf{using-an-alternate-account (1)}

\item\textbf{social-media (1)}

\item\textbf{online-reviews (1)}

\item\textbf{does-not-know-how-to-remove (1)}

\item\textbf{review-what-data-used (1)}

\item\textbf{user-reviews (1)}

\item\textbf{apps-with-access-page-useful (1)}

\item\textbf{location-tracking (1)}

\item\textbf{cost (1)}

\item\textbf{access-to-mobile-device (1)}

\end{itemize}
\clearpage

\end{document}